\documentclass[12pt,preprint]{aastex}

\newcommand{\msun} {$M_\odot$}
\newcommand{\mas} {mas yr$^{-1}$}
\newcommand{\kms} {km s$^{-1}$}
\shorttitle{Moving Objects in the UDF}
\shortauthors{Kilic, Gianninas, \& von Hippel}

\begin{document}

\title{Moving Objects in the Hubble Ultra Deep Field\footnote{Based on
observations made with the NASA/ESA Hubble Space Telescope, obtained from
the Data Archive at the Space Telescope Science Institute, which is operated
by the Association of Universities for Research in Astronomy, Inc., under
NASA contract NAS 5-26555.}}

\author{Mukremin Kilic and Alexandros Gianninas}
\affil{Homer L. Dodge Department of Physics and Astronomy, University of Oklahoma,\\
440 W. Brooks St., Norman OK, 73019, USA}
\author{Ted von Hippel}
\affil{Embry-Riddle Aeronautical University,\\ 600 S. Clyde Morris Blvd., Daytona Beach, FL 32114} 
\email{kilic@ou.edu, alexg@nhn.ou.edu, ted.vonhippel@erau.edu}

\begin{abstract}

We identify proper motion objects in the Hubble Ultra Deep Field (UDF) using the optical data from the
original UDF program in 2004 and the near-infrared data from the 128-orbit UDF 2012 campaign.
There are 12 sources brighter than $I=27$ mag that display $>3\sigma$ significant proper motions. We do not find
any proper motion objects fainter than this magnitude limit. Combining optical and near-infrared photometry,
we model the spectral energy distribution of each point-source using stellar templates and state-of-the-art
white dwarf models. For $I\leq27$ mag, we identify 23 stars with K0-M6 spectral types and two faint blue objects
that are clearly old, thick disk white dwarfs. We measure a thick disk white dwarf space density of 
$0.1-1.7 \times 10^{-3}$ pc$^{-3}$ from these two objects. There are no halo white dwarfs in the UDF down to $I=27$ mag.
Combining the Hubble Deep Field North, South, and the UDF data, we do not see any evidence for dark matter in the
form of faint halo white dwarfs, and the observed population of white dwarfs can be explained with the standard
Galactic models.

\end{abstract}

\keywords{Galaxy: halo, stellar content, structure---stars: Population II---white dwarfs---proper motions}

\section{INTRODUCTION}

The deepest image of the Universe acquired with the {\em Hubble Space Telescope (HST)}, the
UDF \citep{beckwith05}, provides a new opportunity to study the structure of the Galactic
disk and the halo to its limits. Previously, \citet{ibata99} and \citet{mendez00} used the Hubble Deep
Field North \citep{williams96} and South \citep{casertano00} data to constrain the stellar content
of the Galaxy. They proposed that the faint blue objects observed in these fields are old halo white
dwarfs that would be consistent with the observed microlensing events toward the Large Magellanic Cloud
\citep{alcock00} and would explain part of the dark matter in the solar neighborhood. \citet{kilic04,kilic05},
\citet{pirzkal05}, and \citet{mahmud08} showed that some of these faint blue objects
are confused with quasars and an extensive study including proper motion measurements and spectral energy
distribution (SED) fitting is required to identify stars in deep {\em HST} images.

The UDF is the only deep field that is studied spectroscopically by the pioneering work of \citet{pirzkal05}.
Using the Advanced Camera for Surveys (ACS) observations of the UDF and the low resolution
spectroscopy from the Grism ACS program for Extragalactic Science survey \citep[GRAPES,][]{pirzkal04},
\citet{pirzkal05} identified 26 stars and 2 quasars brighter than $I=27$ mag and additional 18 unresolved
sources with $27<I<29.5$ mag. They defined a stellarity index ($S_{\rm i}$) using the curve of
growth analysis of the light distribution of each object, and identified all objects with
$S_{\rm i}\leq0.05$ as unresolved sources. More importantly, the GRAPES spectra enabled \citet{pirzkal05}
to differentiate blue extra-galactic objects, i.e. quasars, from stars down to fainter magnitudes.
They spectroscopically confirmed two of the 18 faint unresolved sources as quasars, and classified the remaining 16 sources
as stars. Assuming that these objects are main-sequence stars, they would have to be at distances larger than 300 kpc.
Therefore, the only viable explanation for these objects would be faint white dwarfs in the thick disk or halo of the
Galaxy. If these are high-velocity white dwarfs, they could contribute $<10$\% to the total dark matter halo mass.
On the other hand, not all of these sources are expected to be stars and some contamination from blue extra-galactic
sources is likely.

\citet{pirzkal05} split the UDF data set into two halves with a baseline of 73 days, and demonstrated
that there are no sources with proper motion $\mu\geq27$ \mas. \citet{mahmud08} used shallower data from
two additional epochs with a 3 yr baseline to identify proper motion objects, finding 7 objects with
significant proper motions.

Here we take advantage of the Wide Field Camera 3 infrared data from the UDF12 campaign, a 128-orbit large
{\em HST} program \citep{ellis13,koekemoer12},
to measure proper motions for the point sources identified by \citet{pirzkal05}, including sources fainter than
$I=27$ mag. These data were obtained in 2012 August-September, and provide an $\approx$8.7 yr baseline for astrometry.
\citet{koekemoer12} describe data reduction and calibration procedures for these observations.
The UDF12 observations in four near-infrared filters, F105W, F125W, F140W, and F160W, also greatly
extend our ability to constrain the spectral type for each object.
The deepest images are in the F105W (100 orbits) and F160W (84 orbits) bands, which
correspond roughly to J and H filters, respectively. The $5\sigma$ limiting sensitivity of the F105W image reaches
an AB magnitude of 30. Hence, these data provide the best opportunity yet to measure proper motions and to model
the SEDs of the stars in the UDF.

Section 2 describes the identification of point sources, while \S 3 presents our proper motion measurements.
\S 4 describes our SED fitting procedures and classifications for the
faint sources in the UDF. We discuss various implications of these results in \S 5.

\section{IDENTIFICATION OF POINT SOURCES}

To identify unresolved sources, \citet{pirzkal05} used the IRAF task RADPROF
and performed aperture photometry using increasing aperture sizes. They
fit a cubic spline to these measurements to create a point spread function (PSF)
for each object. They compared each PSF with the empirical combined-PSF from
bright unresolved objects, and classified objects with PSF distributions similar
to this empirical PSF as unresolved sources. This approach works well for bright targets,
but it is likely to fail for nearly unresolved faint objects.
The discovery of 16 unresolved sources with $I\geq27$ mag in the UDF
is potentially important, but clearly unexpected.

To verify the classification of point sources and to derive precise centroids for each target,
we use the IRAF DAOPHOT package to create a PSF template using bright, unsaturated, isolated targets
in the I-band (deepest) image and use this template to fit each object. Figure 1 shows the sharpness parameter ($S_I$)
derived from DAOPHOT as a function of magnitude. Stars have $S_I\approx0$, whereas
resolved objects have increasingly larger $S_I$ based on their morphology.
This figure demonstrates that the point sources can be identified reliably down to about $I=27$ mag.
\citet{pirzkal05} identified 28 point sources, including two spectroscopically confirmed quasars,
brighter than $I=27$ mag. All but one of these sources, UDF 4322, have $S_I$ indicative of
stars. 

Figure 2 presents the PSF distributions for 23 of these 28 sources that are not saturated in the I-band image.
$S_I$ for each source is also given in each panel.
We use the IRAF task PRADPROF to plot the radial profile of each object. We also plot the I-band PSF template
in each panel for a direct comparison. A comparison of the PSF for each object with our template PSF shows
that all but one of these sources have PSF distributions consistent with being unresolved. The PSF distribution
for UDF 4322 is slightly shallower than the other point sources, and UDF 4322 is likely a resolved object. 
All of the unresolved objects in the UDF, except saturated sources, have $S_I\leq$0.12, while
UDF 4322 has $S_I=0.32$. We note that UDF 4322 also has an SED significantly different from the
stellar objects (see \S 4).

Figure 3 presents the contour maps of the flux distribution around UDF 4322 and an unresolved source
with a similar brightness. A comparison of the contour maps for UDF 4322 ($I=26.84$ mag) with UDF 443 ($I=26.92$ mag)
shows that unresolved sources have circular contour maps, whereas UDF 4322 is elongated, and slightly resolved in the 
$I$ band image. Excluding UDF 4322 from the list of unresolved objects and the two spectroscopically confirmed
quasars (UDF 6732 and 9397), there are 25 sources brighter than $I=27$ mag that are clearly stellar.

The morphological classification of point sources is more problematic for fainter magnitudes. \citet{pirzkal05} classify
18 objects with $I\geq27$ mag as point sources, including two quasars (UDF 4120 and 8157). Figure 4 presents the PSF
distributions of these 18 sources compared to the I-band PSF template derived from brighter stellar sources. It is
clear from this figure that the majority of these fainter sources are likely resolved objects. Only two of these
sources, UDF 7113 and 8081, have radial profiles consistent with unresolved objects and $S_I\approx0$.
The radial profiles for the remaining 16 targets are too shallow to be stellar.

\section{PROPER MOTION MEASUREMENTS}

Proper motion measurements are the best way to identify stars in deep {\em HST} images, and to distinguish between
unresolved quasars and stellar objects. They are also crucial for constraining the kinematic properties of each object,
and assigning membership in the Galactic disk or halo.

To identify high proper motion objects in the UDF, we used the deepest images from the UDF 2004 and UDF12 datasets;
I(F775W)-, J(F105W)-, and H(F160W)-band images. \citet{beckwith05} provided source catalogs for the first epoch
data using the Source Extractor package \citep{bertin96}, which is designed to work best for resolved objects. We used the DAOPHOT package
to create a PSF template for each filter and used this template to precisely constrain the centroids for each object in
each epoch. 

We identified $>$200 compact objects (isolated, low residuals, and not fuzzy) that can be used as reference objects to define
an absolute reference frame. These sources have half light radii R50 $<$ 4 pixels, ellipticity $<$ 0.5, full width
at half maximum $<$ 8 pixels, and stellarity index (as defined by Source Extractor) larger than 0.7.
We visually inspected all of these sources in different filters to avoid any mismatches.
We used the IRAF routine GEOMAP to fit a quadratic polynomial to map the distortions and deleted
deviant points using a $3\sigma$ rejection algorithm. Rejection of very deviant points is required
because the reference objects are compact galaxies and centroiding errors are larger for galaxies.
After mapping the distortions with the GEOMAP package, we transformed the object coordinates
to the second epoch positions with the GEOXYTRAN routine. Using the reference objects, we confirm the
pixel size difference between the optical and infrared images (30 mas versus 60 mas pixel$^{-1}$) and
that their orientations are aligned to better than 0.002$^{\circ}$. 

Figure 5 presents the differences in position between the UDF04 I-band and the UDF12 F105W images for 200 compact
objects that form our reference frame. Red circles mark objects brigher than $I=27$ mag. The majority of
these compact sources do not show any positional differences over the 8.7 yr baseline. However, there are
seven objects brighter than $I=27$ mag that show significant motion. These sources are labeled in the figure and
they are clearly stars. We check these results using our second deepest image in the UDF12 dataset, the F160W
image. The dotted lines mark the location of each source in the F160W image. Since the F160W image mostly consists
of the data from the UDF09 program, the baseline between the I and F160W images are shorter. The observed
locations of the seven moving objects in the F160W image are consistent with the F105W image positions, providing
further evidence that these seven relatively bright objects are clearly moving.

There are several fainter objects with $I>27$ mag that show significant motion in the F105W image. However, none
of these objects show the same motion in the F160W image, indicating that they are likely resolved galaxies.
Color-dependent morphological differences in these galaxies may cause our PSF-fitting algorithm to find
slightly different centroids for these faint sources, and explain the differences in positions measured
from the I-band, F105W, and F160W images. Hence, we do not find any moving objects fainter than $I=27$ mag.

The main source of error in our proper motion measurements is the positions of the reference compact
objects (galaxies). The residuals in the coordinate transformations are 0.3 pixels (9 mas) in each coordinate.
Given the 8.7 year baseline between the UDF04 and UDF12 programs, this corresponds to 1.04 \mas\
errors in each coordinate, or 1.47 \mas\ total proper motion errors for each source.
Tables 1 and 2 present the proper motions for 46 unresolved source candidates identified by \citet{pirzkal05}.
The near-infrared images from the Wide Field Camera 3 cover an area smaller than the optical data. Hence, some
of these sources are not in the UDF12 dataset. On the other hand, a few of the brighter sources have proper motion
measurements from an earlier epoch \citep{mahmud08}. We include these measurements in Table 1 to have a nearly
complete list of proper motions for each source. In total, there are 12 sources with ($>3\sigma$) significant proper motions.

Figure 6 compares proper motion measurements for 11 objects that are common between our study
and that of \citet{mahmud08}. The latter study is limited to $I<27$ mag objects due to shallower data.
Our measurements agree with the \citet{mahmud08} results within $1\sigma$ errors. This
gives us confidence that our proper motion measurements are reliable.

\section{SPECTRAL TYPES OF STARS IN THE UDF}

\subsection{Bright ($I<27$ mag) Unresolved Sources} 
\citet{pirzkal05} determined the spectral types of the unresolved objects in the UDF by fitting stellar templates to
the low resolution grism spectroscopy from the GRAPES survey. The spectroscopic data quality degrades with increasing
magnitude and the background subtraction becomes the limiting factor for the objects fainter than $I=27$ mag.
The SEDs for the majority of the unresolved sources are best-fit with late type K-M stars, where infrared data
would be extremely useful, but was unavailable for the initial analysis by \citet{pirzkal05} and \citet{mahmud08}.

Using the F105W, F125W, F140W, and F160W data in the UDF12 program, we perform PSF-photometry for the unresolved
sources detected in these images. We use an aperture size of $0.4\arcsec$ and AB magnitude zeropoints of
26.0974, 26.0449, 26.2608, and 25.7551 for the F105W, F125W, F140W, and F160W images, respectively. In addition,
a few of the targets outside the field-of-view of the UDF12 observations have infrared photometry available from
NICMOS observations in the F110W and F160W filters \citep{coe06}. Table 3 presents all available optical and infrared
photometry for the 46 unresolved source candidates.

We attempt to fit the SEDs of the UDF objects by combining B, V, I, and z photometry from \citet{beckwith05},
our own photometry in the F105W, F125W, F140W, and F160W filters, and the NICMOS photometry in the F110W and F160W filters.
We use the IRAF task CALCPHOT, which is designed for simulating the HST observations, to simulate the SEDs for main-sequence
stars using \citet{pickles98} stellar templates for O5 to M6 dwarfs. We assign spectral types to each object using a
$\chi^2$ minimization technique, where photometry in each band is weighted according to its error.

Figure 7 shows the optical and infrared SEDs and our best-fit templates for the 26 bright ($I<27$ mag) unresolved 
source candidates, excluding the quasars UDF 6732 and 9397.
Our classifications for the stars agree reasonably well with the \citet{pirzkal05} classifications, but are superior
to the previous analysis due to the combination of optical and infrared photometry. One of these sources, UDF 4322,
has an optical SED similar to A7 type stars. However, its infrared SED shows a flux excess that is similar to
the spectroscopically confirmed quasars in the UDF. Hence, based on its morphology (Fig. 2 and 3) and optical and
infrared SED, UDF 4322 is most likely an extragalactic object. 

The SEDs for the remaining 25 targets fit spectral templates fairly well. We identify 17 M dwarfs, 7 K dwarfs, and
one F type star (UDF 9020), which is certainly a white dwarf (see the discussion below). In addition, UDF 4839
has optical photometry that is consistent with a K2 dwarf, but its infrared photometry is significantly fainter
than expected for K dwarfs. This star must be a white dwarf as well.

One of our targets, UDF 1147, is the brightest object in the $B$ filter and overly saturated in that image.
Its spectrum implies a $\sim$F7 type star, though its $VIz$ photometry is more consistent with a K3 dwarf star.
Therefore, we conclude that either its $B$ photometry is wrong, or it may have a hotter companion.
Unfortunately, the GRAPES spectrum does not go blue enough to confirm this result, but the blue excess is apparent
in its photometry with $B-V=-0.13$ and $V-I=0.49$ mag.

\citet{pirzkal05} identified two L dwarf candidates, UDF 366 and 443. Unfortunately, these objects are not in the field
of view of the NICMOS or WFC3 infrared observations. The optical SEDs for these two sources match an M6 spectral template,
which is the latest spectral type available in the \citet{pickles98} library. Follow-up near-infrared observations will be useful to
constrain the spectral types for UDF 366 and 443 more precisely.

\subsection{Faint ($I>27$ mag) Unresolved Sources}

Figure 8 shows optical and infrared color-color diagrams for the 46 unresolved source candidates along with the synthetic colors for O5-M6
type stars and cool white dwarfs with pure H and pure He atmospheres \citep{tremblay09,bergeron11}. These 46 sources follow the stellar sequence in the optical
color-color diagrams (top panel). However, the majority of the faint ($I>27$ mag) sources and UDF 4322
have infrared colors significantly different than stars (bottom panel). These objects have infrared colors similar to
the spectroscopically confirmed quasars UDF 6732 and 9397.
\citet{pirzkal05} note that some of these faint sources are hard to distinguish from extragalactic sources
without higher signal-to-noise ratio spectra or accurate proper-motion measurements. Contamination from
the large number of extra-galactic objects is inevitable at these faint magnitudes. Our proper motion measurements
for these sources (Table 2), their radial profiles (Fig. 4), and infrared photometry demonstrate that the
majority of the faint unresolved source candidates are extragalactic objects.

Among the 18 faint ($I>27$ mag) unresolved source candidates, none show significant proper motion
(Table 2), two are spectroscopically confirmed quasars, and 14 are most likely resolved objects with radial profiles
shallower than the template PSF (Fig. 4). UDF 7113 and 8081 are the only objects with steep radial profiles similar
to the unresolved stellar objects brighter than $I=27$ mag. Their sharpness parameters from DAOPHOT ($\leq0.06$) are
consistent with unresolved sources. 

Figure 9 shows the optical and infrared SEDs and best-fit stellar templates for UDF 7113 and 8081.
UDF 7113 has colors best-matched by a K7 star, while UDF 8081 has colors that are
similar to spectroscopically confirmed quasars. Even though the optical portion of UDF 8081's SED is best-matched by
an A2 type star, its infrared SED is too red for an A type star or a white dwarf. Hence, UDF 8081 is most likely an extragalactic
object. UDF 9006 is potentially interesting; its sharpness parameter of 0.18 and its radial profile is similar to the unresolved
sources in the UDF. However, UDF 9006 also has an infrared SED that is too red for a blue main-sequence star or a white dwarf.
UDF 7113 is the only target fainter than $I=27$ mag that has a radial profile and an optical and infrared SED
consistent with stars.

\section{DISCUSSION}

\subsection{Late type Stars}

There are 25 stars in the UDF down to $I=27$ mag, including 24 K0 or later type stars and 1 F5 type star.
Fainter than this limit, UDF 7113 ($I=27.67$ mag) is the only source that looks like a star, bringing the total
number of stars in the UDF to 26. To estimate photometric distances, we adopt the absolute magnitude ($M_{\rm V}$)
for each spectral type from \citet{pickles98}, and use it to calculate synthetic absolute magnitudes in the ACS F606W
filter ($M_{\rm F606W}$). We calculated photometric distances for all of the objects in our sample using the synthetic
$M_{\rm F606W}$ and the observed $V_{\rm F606W}$ magnitudes. Spectral types from our SED fitting procedure, proper
motions, absolute magnitudes, distances, and tangential velocities are given in Table 1. The photometric errors for
the UDF point sources are relatively small; the resulting errors in the estimated distances are also very small. However,
if the assigned spectral types are wrong by 1 index, the absolute magnitudes could be wrong by as much as 1 mag.

The stellar templates we use have approximately solar metallicity. Therefore, the distances to the metal
poor halo objects, which will be intrinsically fainter for the same spectral type, are over-estimated by our SED
fitting method. To verify the effect of different metallicites, we use synthetic spectra from a PHOENIX model atmosphere grid
\citep{brott05} for stars with $T_{\rm eff} = 2,000-10,000$ K, $[Fe/H]= 0, -2$, and $\log(g) = 4.5$.
The best fit temperature for the halo metallicity models are on average 300 K cooler (see the last two columns in Table 1).
This difference in temperature implies that the distances and tangential velocities presented in Table 1 are
likely overestimated for metal-poor stars by a factor of $\geq2$ \citep{bressan12}. Despite these potential systematic
errors, available kinematic data is sufficient to identify disk and halo stars and white dwarfs.

The majority of the K0 and later type dwarf stars are consistent with being a late type star in the disk or halo of
the Galaxy. Out of the 25 bright stars, 23 are at distances in the range 1 to 62 kpc. UDF 7113 is consistent with
a K7 dwarf at a distance of $\sim$100 kpc and $V_{\rm tan}=340$ \kms. Even though there are several objects with
relatively large $V_{\rm tan}$, given the uncertainties due to unknown metallicities and the errors in our proper
motion measurements (on the order of 1.4 \mas), these tangential velocity estimates are consistent with halo
membership within the errors.

We used the analytical form of the density profile for the Galactic disk and the halo \citep{gilmore90,vonhippel90}
to calculate the expected number of stars in the UDF. Assuming a local normalization of 0.11 M$_\odot$ pc$^{-3}$
\citep{pham97}, we expect to find 3 thin disk, 6 thick disk, and 20 halo stars in the UDF. \citet{reid93} star count
models predict 4 thin disk, 8 thick disk, and 19 halo stars, whereas the Besan\c{c}on galaxy model predicts 32 stars down to 
$I=27$ mag, including two white dwarfs. Given the small field of view of the UDF and the small number statistics, the
observed number of stars (25-26) and the predictions of star count models (29--32) are in good agreement.

\subsection{White Dwarfs}

\citet{pirzkal05} identified four white dwarf candidates in the UDF, objects 4322, 4839, 7768, and 9020. 
Our analysis shows that one of these objects, UDF 4322, is resolved, and therefore almost certainly an
extra-galactic object. Unfortunately, no proper motion information is available for UDF 4322 to support
this argument. However, its infrared colors and SED are not like any known white dwarf, and they are similar to quasars.

Reduced proper motion diagrams provide a clean method for identifying white dwarfs in large surveys \citep{kilic06, kilic10}.
Figure 10 shows a reduced proper motion diagram for the unresolved source candidates in the UDF along with evolutionary sequences
for disk and halo white dwarfs from the models of \citet{tremblay09}. These models include the collision-induced absorption (CIA) due
to molecular hydrogen \citep{borysow97} and the red-wing of the Ly $\alpha$ opacity \citep{kowalski06}.
The model colors turn to the blue around 4000 K due to the CIA.
There are two sources, UDF 4839 and 9020, with $I<27$ mag that are consistent with thick disk white dwarfs.

As a main-sequence star, UDF 4839 would have a tangential velocity well in excess of 5,000 \kms\ at a distance of 100 kpc.
Similarly, UDF 9020 would have a tangential velocity well in excess of 6,000 \kms\ at a distance of 443 kpc. Obviously, both
of these point sources must be white dwarfs. 
The fourth white dwarf candidate, UDF 7768, is consistent with a K7 type main-sequence star at 32 kpc and its reduced proper motion
and colors are similar to the other main-sequence stars in the UDF. Hence, UDF 7768 is most likely a main-sequence star.
This leaves us with two white dwarf candidates brighter than $I=27$ mag, UDF 4839 and 9020. 

Figure 11 shows the spectral energy distributions and the best-fit pure-H (solid line) and
pure-He (dashed line) white dwarf model atmospheres for UDF 4839 and 9020. Thanks to the infrared photometry
from the UDF12 program and \citet{coe06}, for the first time, we are able to compare the SEDs of
young and old white dwarfs simultaneously and find the best-fit solutions for the UDF white dwarf candidates.

UDF 4839 is best-fit by a 5000 K pure-helium atmosphere white dwarf at a distance of 1.6 kpc and $V_{\rm tan}=90$ \kms.
Its infrared photometry shows a slight flux deficit that may be the signature of the CIA due to molecular hydrogen.
However, the infrared photometry is not precise enough to confirm hydrogen in the atmosphere. The best-fit hydrogen atmosphere
model has $T_{\rm eff}=3250$ K, $d=0.6$ kpc, and $V_{\rm tan}=35$ \kms. Regardless of the composition, UDF 4839 is clearly
a 6-10 Gyr old thick disk white dwarf within 0.6-1.6 kpc of the Sun. 

UDF 9020 is best-fit by a 7000 K pure-hydrogen atmosphere white dwarf model at a distance of 4.5 kpc and $V_{\rm tan}=70$ \kms.
The best-fit pure helium atmosphere model also has $T_{\rm eff}=7000$ K. Regardless of the composition, UDF 9020 is
a $\sim1.5$ Gyr old thick disk white dwarf within 4.5 kpc of the Sun. Unfortunately,
we do not have any information on the masses of these white dwarfs. \citet{bergeron05}
emphasized the importance of precise mass determinations in order
to determine the total stellar ages of the white dwarfs and classify their memberships accordingly.
If UDF 9020 is a $\approx0.5$\msun white dwarf, its total age would be $\sim$12 Gyr, consistent with
the thick disk population. In summary, both UDF 4839 and 9020 have PSFs consistent with unresolved objects,
they display proper motion, and their colors are consistent with relatively old white dwarfs in the Galactic
thick disk.

Based on our proper motion measurements and the reduced proper motion diagram presented in Figure 10, UDF 8081
is the only object in our sample that would be consistent with a white dwarf in the halo. However, its SED
demonstrates that UDF 8081 is clearly not a white dwarf. Therefore, we do not find any white dwarfs
in the UDF with halo kinematics.

\subsubsection{Space Density of White Dwarfs}

\citet{pirzkal05} calculated the white dwarf space density in the UDF using a simple 1/$V_{\rm max}$ analysis
\citep{schmidt68} and found a local density of between $3.5 \times 10^{-5}$ and $1.1 \times 10^{-2}$ stars pc$^{-3}$. 
We perform a similar analysis using the two white dwarfs in the UDF. For a limiting magnitude of
$I=27$ mag and using the best-fit white dwarf models, we estimate a local density of 1.3 $\times$ 10$^{-4}$ pc$^{-3}$ for
the thick disk white dwarfs. If UDF 4839 is a very cool DA (Fig. 11) at a distance of 0.6 kpc, the local density of thick
disk white dwarfs would increase to 1.7 $\times$ 10$^{-3}$ pc$^{-3}$. \citet{sion09} estimate a local white dwarf space density of
$4.9 \pm 0.5 \times 10^{-3}$ pc$^{-3}$ using the 20 pc volume-limited sample. The thick disk white dwarfs in the UDF correspond to
3-35\% of the local space density of white dwarfs. This is in good agreement with the
normalization for thick disk white dwarfs \citep{reid05,rowell11}.

The Besan\c{c}on Galaxy models predict two white dwarfs with $I\leq27$ mag. Both are predicted to be thick disk white dwarfs with
$d=1.4-2.2$ kpc and $V_{\rm tan}=25-45$ \kms. These are consistent with the observed properties of UDF 4839 and 9020, the two white
dwarfs brighter than $I=27$ mag in the UDF. The Besan\c{c}on model predicts two additional white dwarfs with $I=27-29$ mag, including
a halo white dwarf with $V_{\rm tan}=235$ \kms. The only faint ($I>27$ mag) unresolved source in the UDF with a star-like SED is UDF 7113.
The best-fit stellar template for UDF 7113 is a K7 main-sequence star. If UDF 7113 is a white dwarf, its SED would be best-matched by a
3750 K pure helium-atmosphere white dwarf at a distance of 2.1 kpc and $V_{\rm tan}=6$ \kms. Obviously, UDF 7113 is not a halo white
dwarf. Finding zero halo white dwarfs, when the expected number is one, is not surprising.
Hence, this analysis shows that the observed number of white dwarfs in the UDF does not require additions to
the standard Galactic models. 

\section{CONCLUSIONS}

A careful analysis of the radial profiles, proper motions, and the SEDs of point-like
sources in the UDF revealed 25 stars brighter than $I=27$ mag and one more likely star (UDF 7113) with $I=27.67$ mag.
Combining the optical photometry with the 1.0-1.6 $\mu$m photometry from the UDF12 program enabled us to fit the SEDs of
each star and constrain their spectral types, distances, and tangential velocities.
Out of the 26 stars, 12 have significant ($>3\sigma$) proper motions, and 24 are consistent with late-type K-M dwarfs
in the disk or halo of the Galaxy. This analysis revealed two stars that are too faint to be main-sequence stars, UDF 4839 and 9020.
These are clearly thick disk white dwarfs at distances of 0.6-4.5 kpc and $V_{\rm tan}=35-90$ \kms. 

All three deep fields observed by the {\em HST}, Hubble Deep Field North, South, and the UDF, have
two white dwarf candidates brighter than $I=27$ mag \citep[][and this study]{kilic04,kilic05}. A comparison of the observed number
of white dwarfs in these deep pencil-beam surveys with model predictions shows that the observed white dwarf population
in the {\em HST} deep fields do not require any additions to the standard Galactic models.

\acknowledgements

\begin{deluxetable}{rrrcrrrcc}
\tabletypesize{\footnotesize}
\tablecolumns{9}
\tablewidth{0pt}
\tablecaption{Physical properties of the bright ($I<27$ mag) unresolved sources}
\tablehead{ \colhead{Object} & \colhead{$\mu$}& \colhead{$\mu_{\rm 2008}$}& \colhead{Spectral}& \colhead{$M_{\rm F606W}$}& \colhead{d}& \colhead{$V_{\rm tan}$} & \colhead{$T_{\rm eff_0}$} & \colhead{$T_{\rm eff_2}$}\\
 & \colhead{(\mas)} & \colhead{(\mas)} & \colhead{Type} & \colhead{(mag)} & \colhead{(kpc)} & \colhead{(\kms)} & \colhead{(K)} & \colhead{(K)}
}
\startdata
19   & \nodata         & \nodata          & M1 &  8.7 &  15  & \nodata & 3800 & 3400 \\ 
366  & \nodata         &  2.42 $\pm$ 1.47 & M6 & 13.6 &   7  & 80 & 2300 & 2000 \\   
443  & \nodata         & \nodata          & M6 & 13.6 &  19  & \nodata & 2200 & 2000 \\ 
834  & \nodata         &  0.27 $\pm$ 0.96 & M5 & 12.2 &   4  & 10 & 2900 & 2700 \\      
911  & \nodata         & 27.01 $\pm$ 0.78 & M4 & 11.1 &   3  & 320 & 3100 & 2800 \\ 
1147 & \nodata         &  5.49 $\pm$ 0.63 & K3 &  6.6 &   4  & 110 & 4800 & 4900 \\ 
2150 & \nodata         & 25.16 $\pm$ 0.45 & M3 & 10.2 &   1  & 140 & 3200 & 2900 \\          
2457 & 7.25 $\pm$ 1.47 &  7.87 $\pm$ 0.82 & M0 &  8.3 &  22  & 750 & 3900 & 3500 \\          
3166 & 7.94 $\pm$ 1.47 &  8.84 $\pm$ 0.66 & K7 &  7.9 &   2  & 90 & 4000 & 3700 \\           
3561 & 2.90 $\pm$ 1.47 & \nodata          & M0 &  8.3 &  62  & 850 & 3900 & 3600 \\          
3794 & 3.09 $\pm$ 1.47 &  4.90 $\pm$ 2.75 & M1 &  8.7 &  41  & 590 & 3900 & 3500 \\           
4322 & \nodata         & \nodata          & A7 &  2.1 & 877  & \nodata & 7600 & 7600 \\           
4839 & \nodata         & 11.64 $\pm$ 1.92 & K2 &  6.0 & 100  & 5530 & 4800 & 4800 \\                 
4945 & \nodata         & \nodata          & M4 & 11.1 &   2  & \nodata & 3100 & 2800\\            
5441 & 7.74 $\pm$ 1.47 &  8.35 $\pm$ 1.14 & M2 &  9.2 &  24  & 860 & 3600 & 3200 \\                 
5921 & \nodata         &  3.27 $\pm$ 1.26 & K7 &  7.9 &   2  & 40 & 4100 & 3800 \\                   
5992 & 1.64 $\pm$ 1.47 &  1.18 $\pm$ 1.71 & K7 &  7.9 &  40  & 310 & 4200 & 4000 \\                 
6461 & \nodata         & \nodata          & M0 &  8.3 &  22  & \nodata & 3900 & 3500 \\            
6732 & 0.13 $\pm$ 1.47 &  0.42 $\pm$ 1.14 & QSO & \nodata & \nodata & \nodata & \nodata & \nodata \\  
7525 & \nodata         &  4.10 $\pm$ 1.50 & M1 &  8.7 &  35  & 680 & 3800 & 3400 \\                
7768 & 5.92 $\pm$ 1.47 &  6.50 $\pm$ 1.14 & K7 &  7.9 &  32  & 910 & 4200 & 4000 \\                
9020 & 3.33 $\pm$ 1.47 &  \nodata         & F5 &  3.5 & 443  & 6980& 6800 & 7000 \\                  
9212 & 11.73 $\pm$ 1.47 & 12.35 $\pm$ 0.54 & M0 &  8.3 & 12  & 650 & 3900 & 3500 \\                   
9230 & 8.99 $\pm$ 1.47 & 10.26 $\pm$ 0.41 & K0 &  5.4 &  10  & 420 & 4900 & 5000 \\                   
9331 & 6.08 $\pm$ 1.47 &  6.39 $\pm$ 0.73 & M4 & 11.1 &   8  & 240 & 3000 & 2800 \\                   
9351 & \nodata         & 12.11 $\pm$ 0.68 & M2.5 & 9.7 &  8  & 440 & 3500 & 3100 \\                    
9397 & 0.47 $\pm$ 1.47 &  0.70 $\pm$ 0.47 & QSO & \nodata & \nodata & \nodata & \nodata & \nodata \\        
9959 & \nodata         & \nodata          & M0  &  8.3 & 47  & \nodata & 4000 & 3600 \\            
\enddata
\tablecomments{$\mu$ and $\mu_{\rm 2008}$ are proper motions from this study and \citet{mahmud08}, respectively. 
$T_{\rm eff_0}$ and $T_{\rm eff_2}$ are the temperatures of the best-fit PHOENIX models with [Fe/H] = 0 and [Fe/H] = $-2$,
respectively.}
\end{deluxetable}

\begin{deluxetable}{cc}
\tabletypesize{\footnotesize}
\tablecolumns{2}
\tablewidth{0pt}
\tablecaption{Proper motion measurements for the faint ($I>27$ mag) unresolved sources}
\tablehead{ \colhead{Object}&
\colhead{$\mu$ (\mas)}
}
\startdata
1343 & \nodata \\
2368 & 2.11 \\
2977 & \nodata \\
3940 & 0.18 \\
4120 & \nodata \\ 
4643 & \nodata \\
5317 & \nodata \\
6334 & 1.25 \\
6442 & 1.01 \\
6620 & 0.33 \\
7113 & 0.58 \\
7194 & 0.65 \\
7357 & 0.44 \\
7894 & 0.29 \\
8081 & 2.36 \\
8157 & 0.56 \\ 
8186 & \nodata \\
9006 & 0.36 \\
\enddata
\tablecomments{UDF 4120 and 8157 are spectroscopically confirmed quasars. All proper motion
measurements are from this study and have errors of 1.47 \mas.}
\end{deluxetable}

\begin{deluxetable}{rcccccccccc}
\rotate
\tabletypesize{\tiny}
\tablecolumns{11}
\tablewidth{0pt} \tablecaption{Optical and infrared photometry (in the AB system) for the unresolved source candidates in the UDF}
\tablehead{ \colhead{Object}&
\colhead{F435W}&
\colhead{F606W}&
\colhead{F775W}&
\colhead{F850LP}&
\colhead{F105W}&
\colhead{F125W}&
\colhead{F140W}&
\colhead{F160W}&
\colhead{F110W}&
\colhead{F160W}
}
\startdata
19   & 26.563$\pm$0.030 & 24.643$\pm$0.004 & 23.626$\pm$0.002 & 23.309$\pm$0.002 & \nodata & \nodata & \nodata & \nodata & \nodata & \nodata \\
366  & 30.731$\pm$1.018 & 27.723$\pm$0.044 & 24.704$\pm$0.003 & 23.423$\pm$0.002 & \nodata & \nodata & \nodata & \nodata & \nodata & \nodata \\
443  & \nodata          & 29.973$\pm$0.200 & 26.917$\pm$0.014 & 25.448$\pm$0.006 & \nodata & \nodata & \nodata & \nodata & \nodata & \nodata \\
834  & 27.416$\pm$0.079 & 25.252$\pm$0.008 & 23.277$\pm$0.001 & 22.496$\pm$0.001 & \nodata & \nodata & \nodata & \nodata & 22.061$\pm$0.072 & 22.029$\pm$0.082 \\
911  & 25.047$\pm$0.017 & 23.053$\pm$0.002 & 21.275$\pm$0.000 & 20.607$\pm$0.000 & \nodata & \nodata & \nodata & \nodata & \nodata & \nodata \\
1147 & 19.530$\pm$0.000 & 19.662$\pm$0.000 & 19.168$\pm$0.000 & 19.040$\pm$0.000 & \nodata & \nodata & \nodata & \nodata & \nodata & \nodata \\
1343 & 27.455$\pm$0.027 & 27.655$\pm$0.021 & 27.791$\pm$0.027 & 27.903$\pm$0.052 & \nodata & \nodata & \nodata & \nodata & \nodata & \nodata \\
2150 & 22.535$\pm$0.003 & 20.558$\pm$0.000 & 19.126$\pm$0.000 & 18.320$\pm$0.000 & \nodata & \nodata & \nodata & \nodata & 18.266$\pm$0.052 & 18.024$\pm$0.064 \\
2368 & 29.250$\pm$0.128 & 27.961$\pm$0.027 & 27.805$\pm$0.026 & 27.972$\pm$0.053 & 27.922$\pm$0.049 & 28.004$\pm$0.064 & 28.071$\pm$0.036 & 28.031$\pm$0.060 & 27.920$\pm$0.317 & 27.522$\pm$0.294 \\
2457 & 26.793$\pm$0.036 & 25.004$\pm$0.005 & 24.047$\pm$0.002 & 23.733$\pm$0.003 & 23.521$\pm$0.012 & 23.377$\pm$0.018 & 23.336$\pm$0.012 & 23.303$\pm$0.007 & 23.578$\pm$0.086 & 23.334$\pm$0.098 \\
2977 & 28.024$\pm$0.040 & 28.160$\pm$0.032 & 28.261$\pm$0.038 & 28.196$\pm$0.062 & \nodata & \nodata & \nodata & \nodata & 28.016$\pm$0.368 & 28.070$\pm$0.436 \\
3166 & 20.753$\pm$0.001 & 19.786$\pm$0.000 & 18.998$\pm$0.000 & 18.628$\pm$0.000 & 18.548$\pm$0.002 & 18.438$\pm$0.003 & 18.362$\pm$0.001 & 18.281$\pm$0.001 & 18.697$\pm$0.052 & 18.497$\pm$0.064 \\
3561 & 29.013$\pm$0.121 & 27.263$\pm$0.017 & 26.360$\pm$0.008 & 26.119$\pm$0.012 & 25.745$\pm$0.014 & 25.612$\pm$0.008 & 25.574$\pm$0.010 & 25.570$\pm$0.008 & 25.986$\pm$0.134 & 25.780$\pm$0.137 \\
3794 & 28.869$\pm$0.126 & 26.740$\pm$0.013 & 25.757$\pm$0.006 & 25.477$\pm$0.008 & 25.143$\pm$0.011 & 25.012$\pm$0.015 & 24.996$\pm$0.009 & 24.955$\pm$0.008 & 25.251$\pm$0.111 & 25.232$\pm$0.116 \\
3940 & 27.084$\pm$0.017 & 27.562$\pm$0.019 & 27.704$\pm$0.024 & 27.917$\pm$0.051 & 27.652$\pm$0.023 & 27.599$\pm$0.026 & 27.682$\pm$0.021 & 27.332$\pm$0.013 & 27.319$\pm$0.202 & 26.639$\pm$0.166 \\
4120 & 28.143$\pm$0.045 & 28.046$\pm$0.029 & 27.876$\pm$0.028 & 26.631$\pm$0.016 & \nodata & \nodata & \nodata & \nodata & 27.212$\pm$0.229 & 27.521$\pm$0.311 \\
4322 & 26.856$\pm$0.021 & 26.816$\pm$0.014 & 26.843$\pm$0.016 & 26.913$\pm$0.030 & \nodata & \nodata & \nodata & \nodata & 26.831$\pm$0.194 & 26.017$\pm$0.169 \\
4643 & 32.672$\pm$2.017 & 30.054$\pm$0.130 & 29.259$\pm$0.072 & 29.025$\pm$0.102 & \nodata & \nodata & \nodata & \nodata & 28.710$\pm$0.667 & 99.000$\pm$28.498 \\
4839 & 27.232$\pm$0.030 & 26.007$\pm$0.007 & 25.594$\pm$0.005 & 25.523$\pm$0.009 & \nodata & \nodata & \nodata & \nodata & 25.537$\pm$0.125 & 26.120$\pm$0.147 \\
4945 & 24.752$\pm$0.013 & 22.925$\pm$0.002 & 21.215$\pm$0.000 & 20.536$\pm$0.000 & \nodata & \nodata & \nodata & \nodata & \nodata & \nodata \\
5317 & 28.915$\pm$0.078 & 28.625$\pm$0.041 & 28.584$\pm$0.044 & 28.838$\pm$0.097 & \nodata & \nodata & \nodata & \nodata & \nodata & \nodata \\
5441 & 27.817$\pm$0.079 & 26.052$\pm$0.011 & 24.781$\pm$0.004 & 24.384$\pm$0.005 & 24.040$\pm$0.006 & 23.897$\pm$0.007 & 23.881$\pm$0.005 & 23.867$\pm$0.005 & 24.137$\pm$0.094 & 24.039$\pm$0.106 \\
5921 & 20.520$\pm$0.001 & 19.737$\pm$0.000 & 19.072$\pm$0.000 & 18.571$\pm$0.000 & \nodata & \nodata & \nodata & \nodata & \nodata & \nodata \\
5992 & 27.491$\pm$0.041 & 25.888$\pm$0.007 & 25.186$\pm$0.004 & 25.030$\pm$0.006 & 24.736$\pm$0.010 & 24.637$\pm$0.015 & 24.591$\pm$0.013 & 24.545$\pm$0.012 & 25.023$\pm$0.104 & 24.858$\pm$0.115 \\
6334 & 27.789$\pm$0.032 & 27.983$\pm$0.027 & 28.091$\pm$0.032 & 28.197$\pm$0.062 & 28.010$\pm$0.025 & 27.978$\pm$0.039 & 27.661$\pm$0.039 & 27.575$\pm$0.029 & 27.722$\pm$0.382 & 27.638$\pm$0.397 \\
6442 & 28.158$\pm$0.042 & 28.371$\pm$0.036 & 28.356$\pm$0.039 & 28.382$\pm$0.070 & 27.746$\pm$0.022 & 27.628$\pm$0.028 & 27.894$\pm$0.037 & 27.656$\pm$0.024 & 28.502$\pm$0.665 & 28.375$\pm$0.715 \\
6461 & 26.660$\pm$0.033 & 24.990$\pm$0.005 & 24.003$\pm$0.002 & 23.698$\pm$0.003 & \nodata & \nodata & \nodata & \nodata & \nodata & \nodata \\
6620 & 30.811$\pm$0.512 & 28.119$\pm$0.031 & 27.825$\pm$0.026 & 28.020$\pm$0.054 & 27.650$\pm$0.037 & 27.642$\pm$0.027 & 27.650$\pm$0.038 & 27.505$\pm$0.034 & \nodata & 27.063$\pm$0.291 \\
6732 & 25.706$\pm$0.012 & 24.588$\pm$0.003 & 24.626$\pm$0.003 & 24.606$\pm$0.006 & 23.949$\pm$0.020 & 23.872$\pm$0.020 & 23.685$\pm$0.020 & 23.712$\pm$0.022 & 24.238$\pm$0.088 & 23.992$\pm$0.099 \\
7113 & 30.667$\pm$0.518 & 28.376$\pm$0.046 & 27.673$\pm$0.026 & 27.455$\pm$0.037 & 27.145$\pm$0.013 & 27.064$\pm$0.017 & 26.949$\pm$0.013 & 26.898$\pm$0.011 & 27.016$\pm$0.220 & 26.923$\pm$0.231 \\
7194 & 26.921$\pm$0.018 & 27.050$\pm$0.014 & 27.202$\pm$0.018 & 27.243$\pm$0.034 & 27.109$\pm$0.035 & 26.806$\pm$0.034 & 26.953$\pm$0.035 & 26.977$\pm$0.034 & \nodata & \nodata \\
7357 & 27.984$\pm$0.036 & 27.862$\pm$0.023 & 28.055$\pm$0.030 & 28.288$\pm$0.064 & 28.068$\pm$0.021 & 28.115$\pm$0.036 & 28.081$\pm$0.020 & 28.060$\pm$0.032 & 27.801$\pm$0.326 & 27.624$\pm$0.350 \\
7525 & 28.498$\pm$0.113 & 26.416$\pm$0.012 & 25.350$\pm$0.005 & 25.045$\pm$0.007 & \nodata & \nodata & \nodata & \nodata & \nodata & \nodata \\
7768 & 27.119$\pm$0.033 & 25.444$\pm$0.005 & 24.758$\pm$0.003 & 24.553$\pm$0.005 & 24.278$\pm$0.022 & 24.203$\pm$0.021 & 24.169$\pm$0.016 & 24.196$\pm$0.013 & \nodata & \nodata \\
7894 & 27.531$\pm$0.027 & 27.455$\pm$0.018 & 27.589$\pm$0.023 & 27.794$\pm$0.050 & 27.636$\pm$0.031 & 27.486$\pm$0.034 & 27.503$\pm$0.025 & 27.457$\pm$0.031 & \nodata & \nodata \\
8081 & 27.921$\pm$0.032 & 28.097$\pm$0.027 & 28.382$\pm$0.039 & 28.448$\pm$0.072 & 27.952$\pm$0.046 & 28.163$\pm$0.031 & 27.792$\pm$0.037 & 27.612$\pm$0.029 & 28.022$\pm$0.572 & 28.033$\pm$0.668 \\
8157 & \nodata          & 28.728$\pm$0.050 & 28.344$\pm$0.039 & 28.313$\pm$0.067 & 28.310$\pm$0.052 & 28.338$\pm$0.069 & 28.332$\pm$0.067 & 28.353$\pm$0.077 & 28.565$\pm$0.524 & 28.680$\pm$0.667 \\
8186 & 27.974$\pm$0.031 & 28.121$\pm$0.025 & 28.377$\pm$0.037 & 28.646$\pm$0.083 & \nodata & \nodata & \nodata & \nodata & \nodata & \nodata \\
9006 & 27.454$\pm$0.025 & 27.551$\pm$0.019 & 27.538$\pm$0.022 & 27.347$\pm$0.032 & 26.670$\pm$0.027 & 26.727$\pm$0.025 & 27.092$\pm$0.032 & 26.864$\pm$0.023 & \nodata & \nodata \\
9020 & 27.024$\pm$0.018 & 26.731$\pm$0.010 & 26.732$\pm$0.011 & 26.905$\pm$0.022 & 26.741$\pm$0.014 & 26.907$\pm$0.018 & 27.033$\pm$0.015 & 27.123$\pm$0.011 & 26.638$\pm$0.170 & 26.985$\pm$0.209 \\
9212 & 25.494$\pm$0.014 & 23.642$\pm$0.002 & 22.676$\pm$0.001 & 22.373$\pm$0.001 & 22.122$\pm$0.012 & 21.998$\pm$0.026 & 21.971$\pm$0.014 & 21.953$\pm$0.014 & \nodata & \nodata \\
9230 & 21.125$\pm$0.001 & 20.380$\pm$0.000 & 19.999$\pm$0.000 & 19.888$\pm$0.000 & 19.993$\pm$0.014 & 19.953$\pm$0.022 & 19.948$\pm$0.025 & 19.933$\pm$0.009 & 19.866$\pm$0.052 & 19.889$\pm$0.064 \\
9331 & 27.937$\pm$0.093 & 25.706$\pm$0.009 & 23.879$\pm$0.002 & 23.180$\pm$0.002 & 22.746$\pm$0.014 & 22.553$\pm$0.023 & 22.539$\pm$0.017 & 22.536$\pm$0.014 & 23.117$\pm$0.086 & 22.915$\pm$0.096 \\
9351 & 26.101$\pm$0.021 & 24.138$\pm$0.003 & 22.826$\pm$0.001 & 22.386$\pm$0.001 & \nodata & \nodata & \nodata & \nodata & \nodata & \nodata \\
9397 & 21.394$\pm$0.001 & 21.137$\pm$0.000 & 21.020$\pm$0.000 & 20.977$\pm$0.001 & 21.047$\pm$0.020 & 20.940$\pm$0.020 & 20.886$\pm$0.020 & 20.600$\pm$0.022 & 20.738$\pm$0.053 & 20.294$\pm$0.064 \\
9959 & 28.731$\pm$0.140 & 26.676$\pm$0.016 & 25.792$\pm$0.008 & 25.519$\pm$0.010 & \nodata & \nodata & \nodata & \nodata & \nodata & \nodata 
\enddata
\tablecomments{Optical photometry is from \citet{beckwith05} and the last two columns are NICMOS F110W and F160W photometry from \citet{coe06}.}
\end{deluxetable}

\begin{figure}
\includegraphics[width=5in,angle=-90]{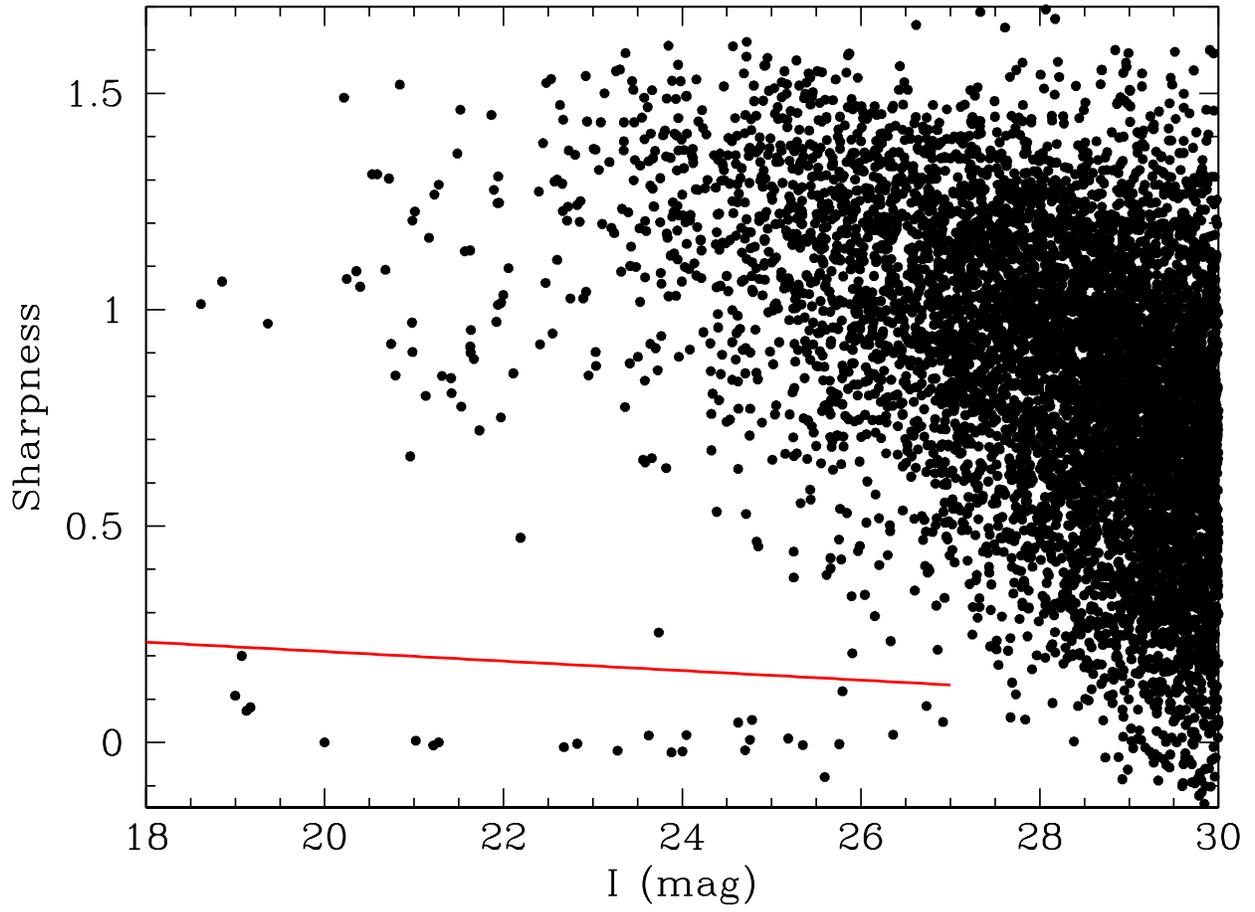}
\caption{The distribution of the sharpness parameter from DAOPHOT for the UDF objects. Objects below the solid line are
unresolved.}
\end{figure}

\begin{figure}
\includegraphics[width=5in,angle=-90]{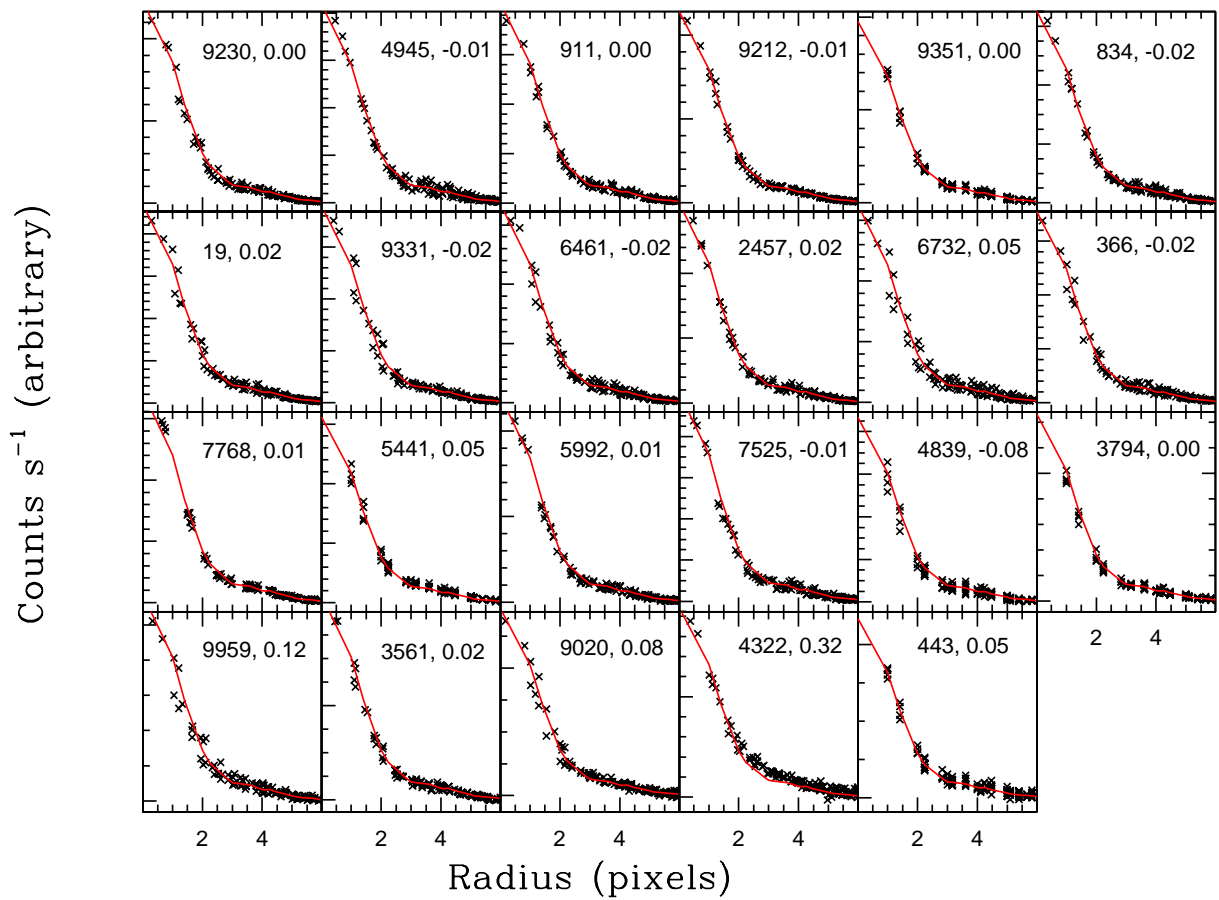}
\caption{Radial profiles of the relatively bright ($I<27$ mag), unsaturated, and unresolved source candidates identified by
\citet{pirzkal05}. The objects are shown in increasing I-band magnitude (from top left to bottom right).
The PSF template derived from bright isolated stars (solid line) is shown in
each panel for comparison. The sharpness parameter for each object is also given after the object name.
UDF 4322 has a shallower profile compared to the PSF template and is likely resolved.}
\end{figure}

\begin{figure}
\includegraphics[width=5in]{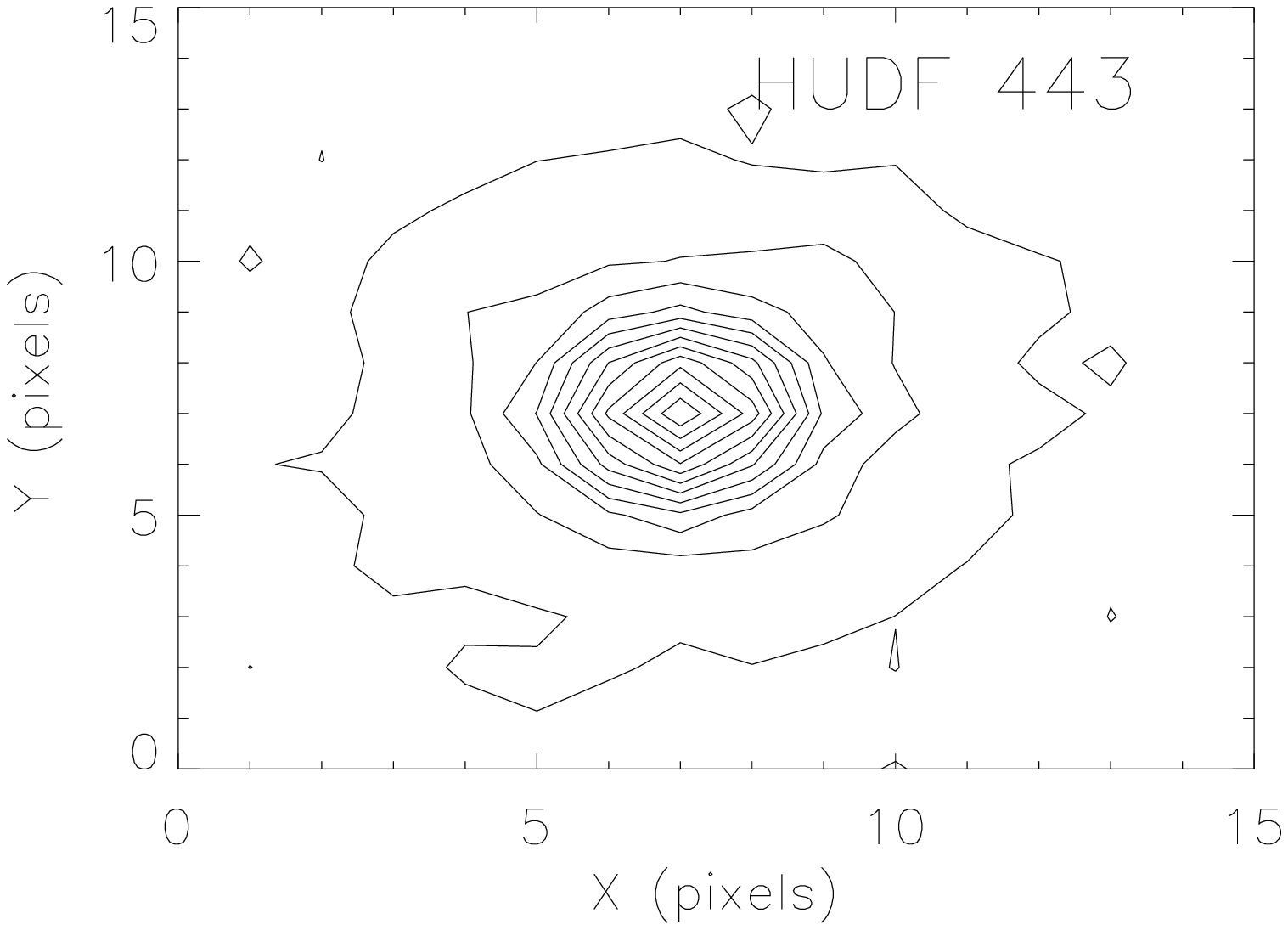}
\includegraphics[width=5in]{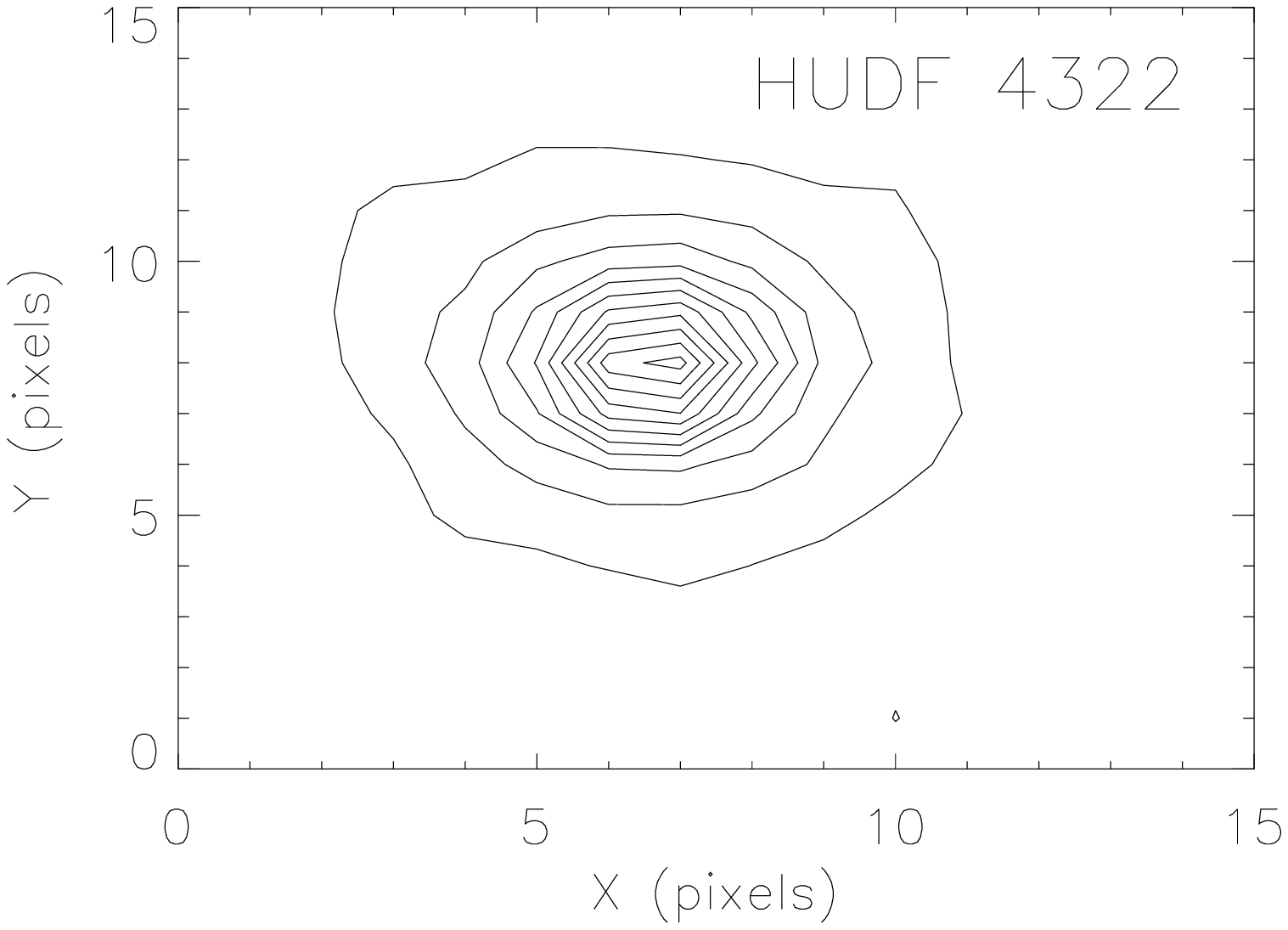}
\caption{Contour maps of the flux distribution around UDF 443 and 4322 in the I-band.}
\end{figure}

\begin{figure}
\includegraphics[width=5in,angle=-90]{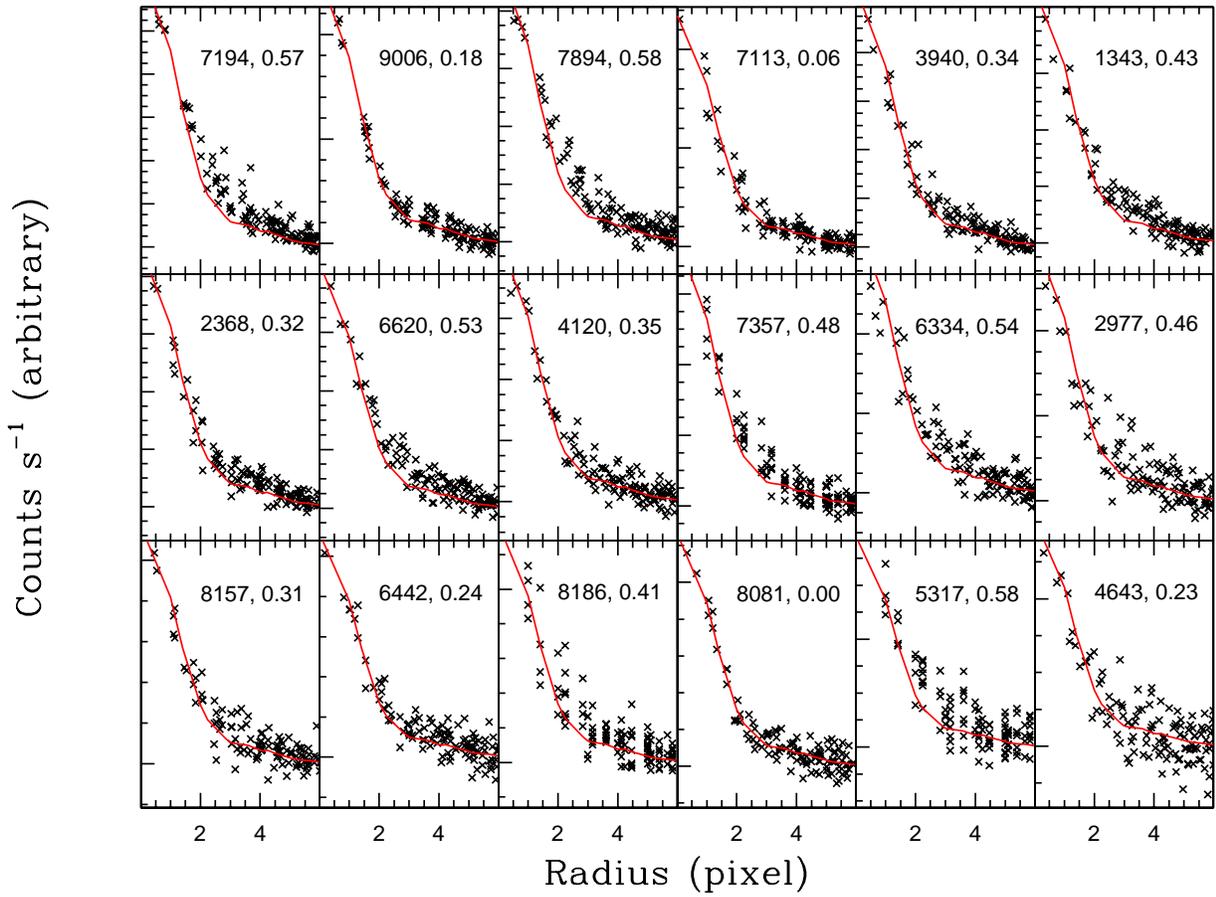}
\caption{Similar to Figure 2, but for faint ($I>27$ mag) unresolved source candidates.
UDF 7113 and 8081 are the only faint objects with sharpness $\approx$0 and radial profiles consistent
with unresolved sources.}
\end{figure}

\begin{figure}
\plotone{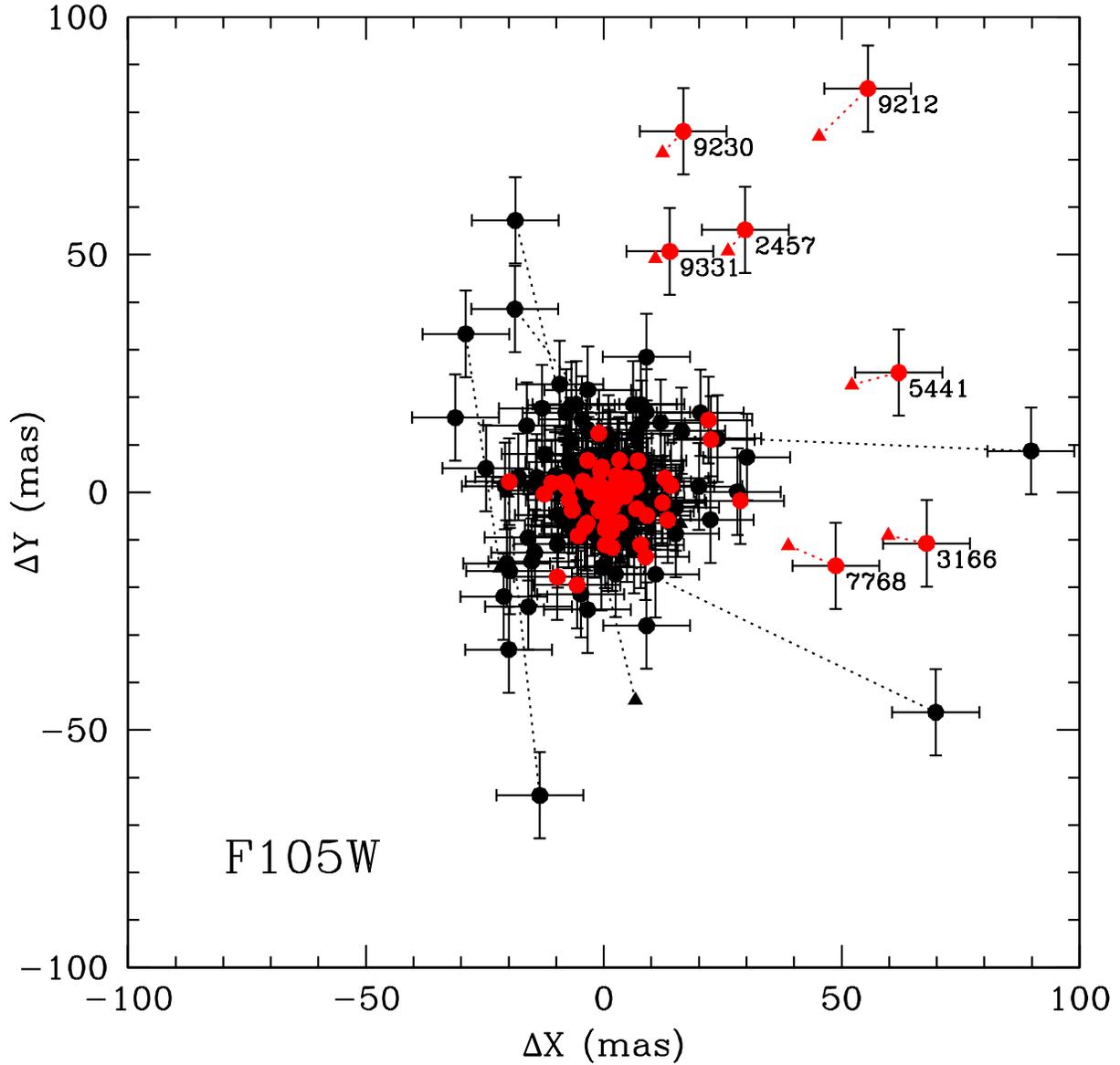}
\caption{Differences in positions for 200 compact UDF sources between the 2004 I-band and 2012 F105W images (circles). 
Triangles represent the positions of the same sources in the F160W image. The dotted lines connect the astrometric data
derived from the F105W and F160W filters. Objects brighter than $I=27$ mag
are marked by red points and the objects with significant motion are labeled.}
\end{figure}

\begin{figure}
\plotone{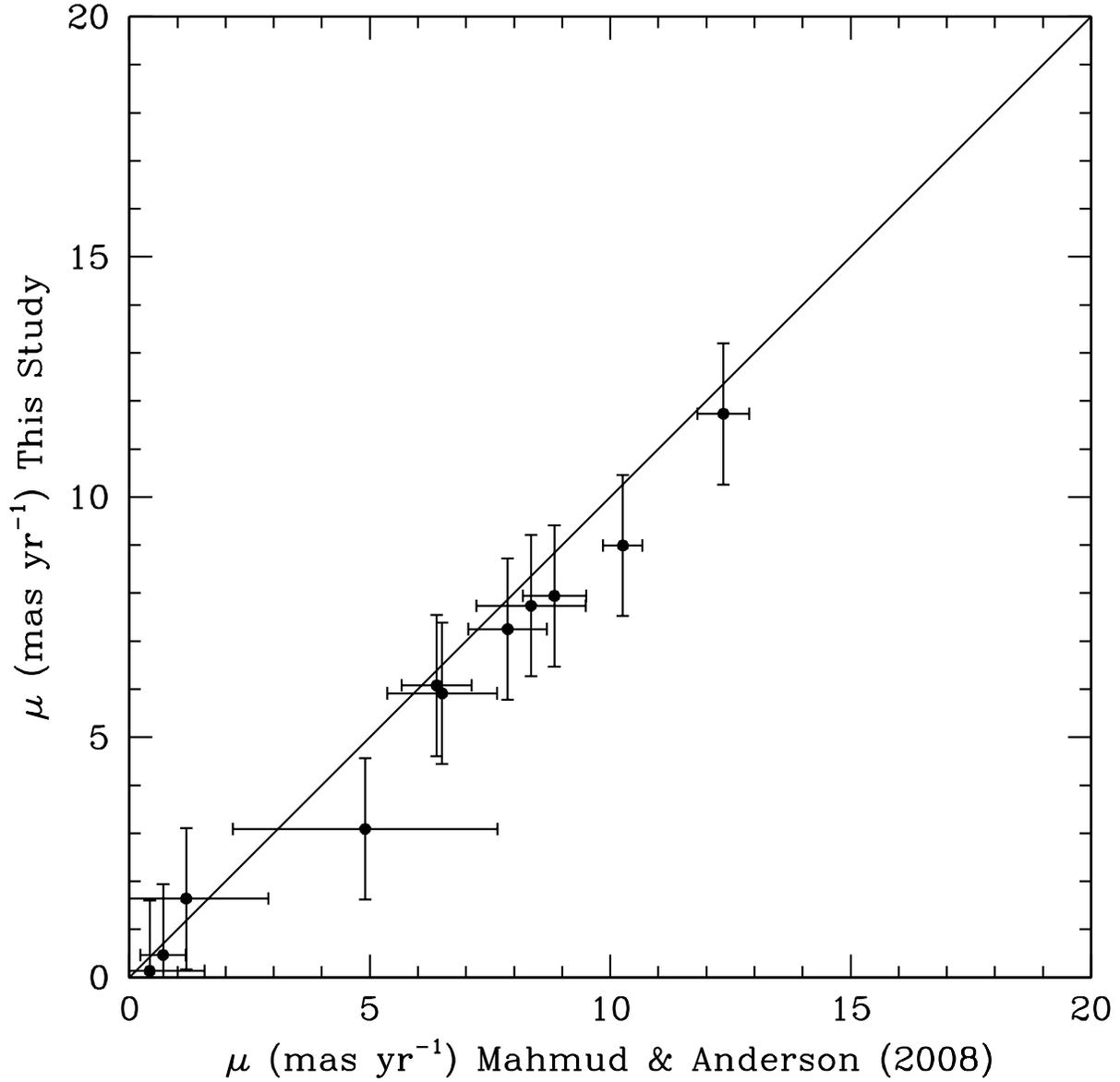}
\caption{A comparison between proper motion measurements for 11 objects that are common between this study
and \citet{mahmud08}. The results agree within $1\sigma$ errors.}
\end{figure}

\begin{figure}
\includegraphics[width=4in,angle=-90]{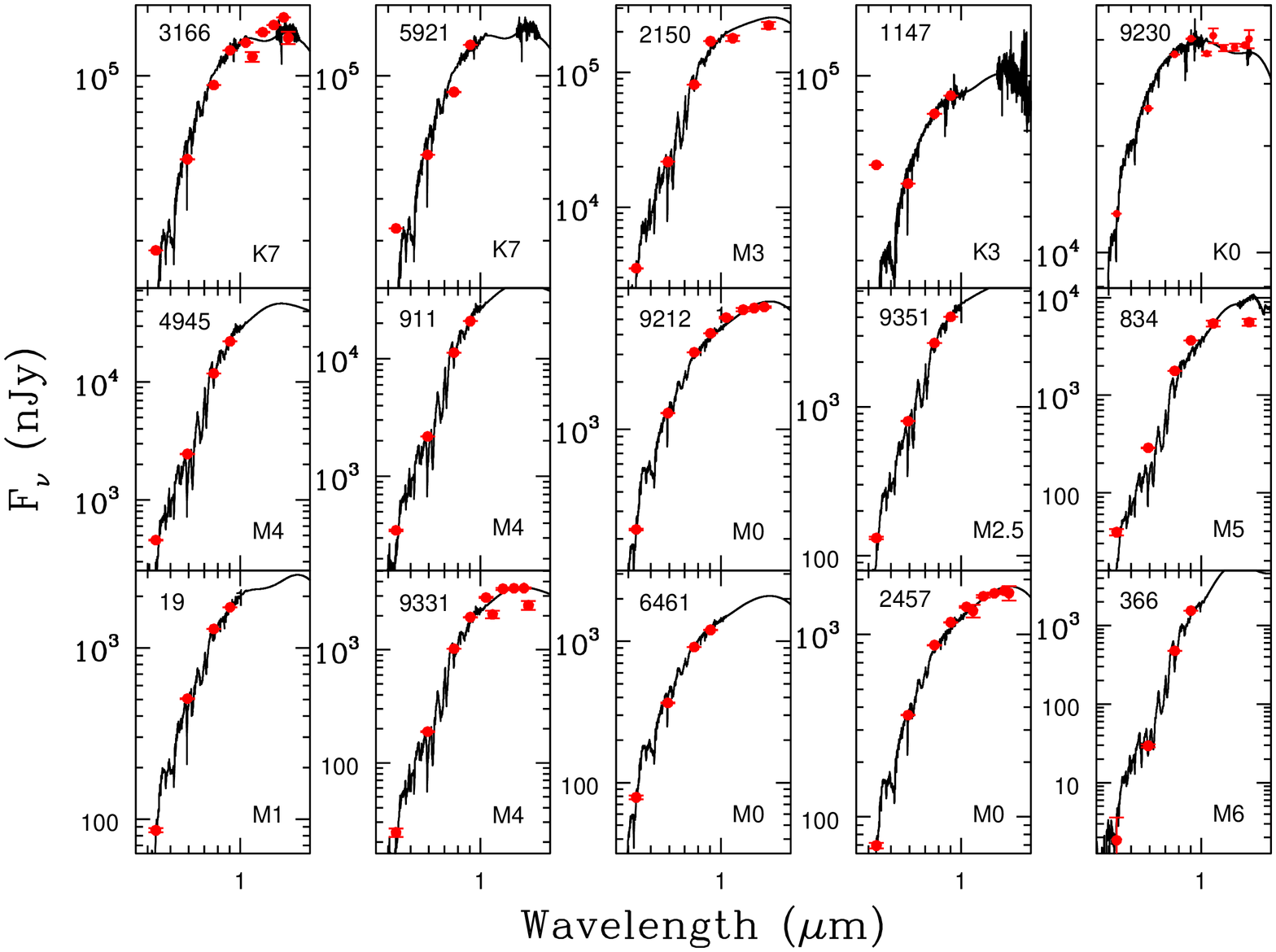}
\includegraphics[width=4in,angle=-90]{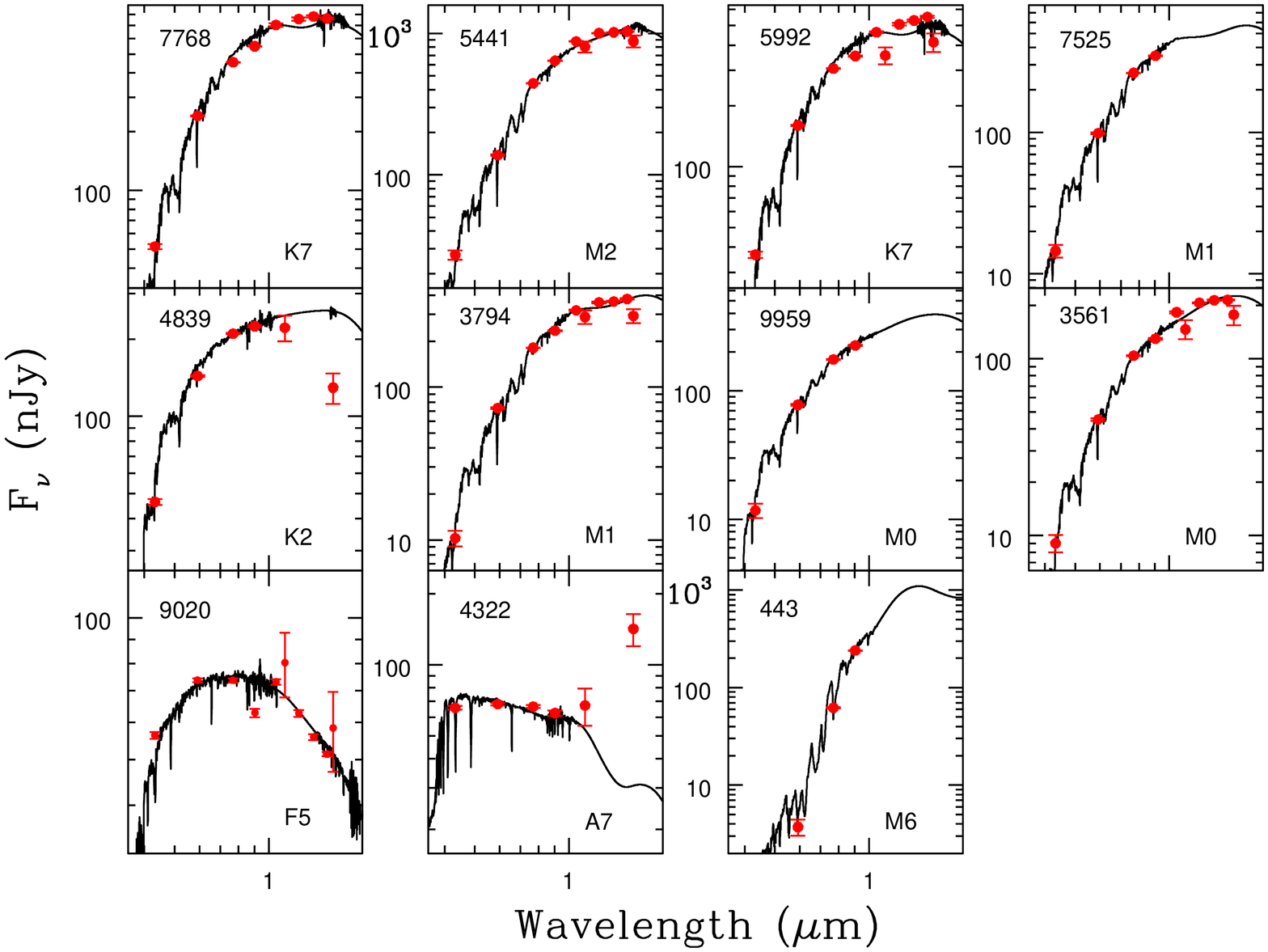}
\caption{Optical and infrared SEDs and best-fitting main-sequence star templates \citep{pickles98} for the bright
($I<27$ mag) unresolved source candidates, excluding quasars.}
\end{figure}

\begin{figure}
\plotone{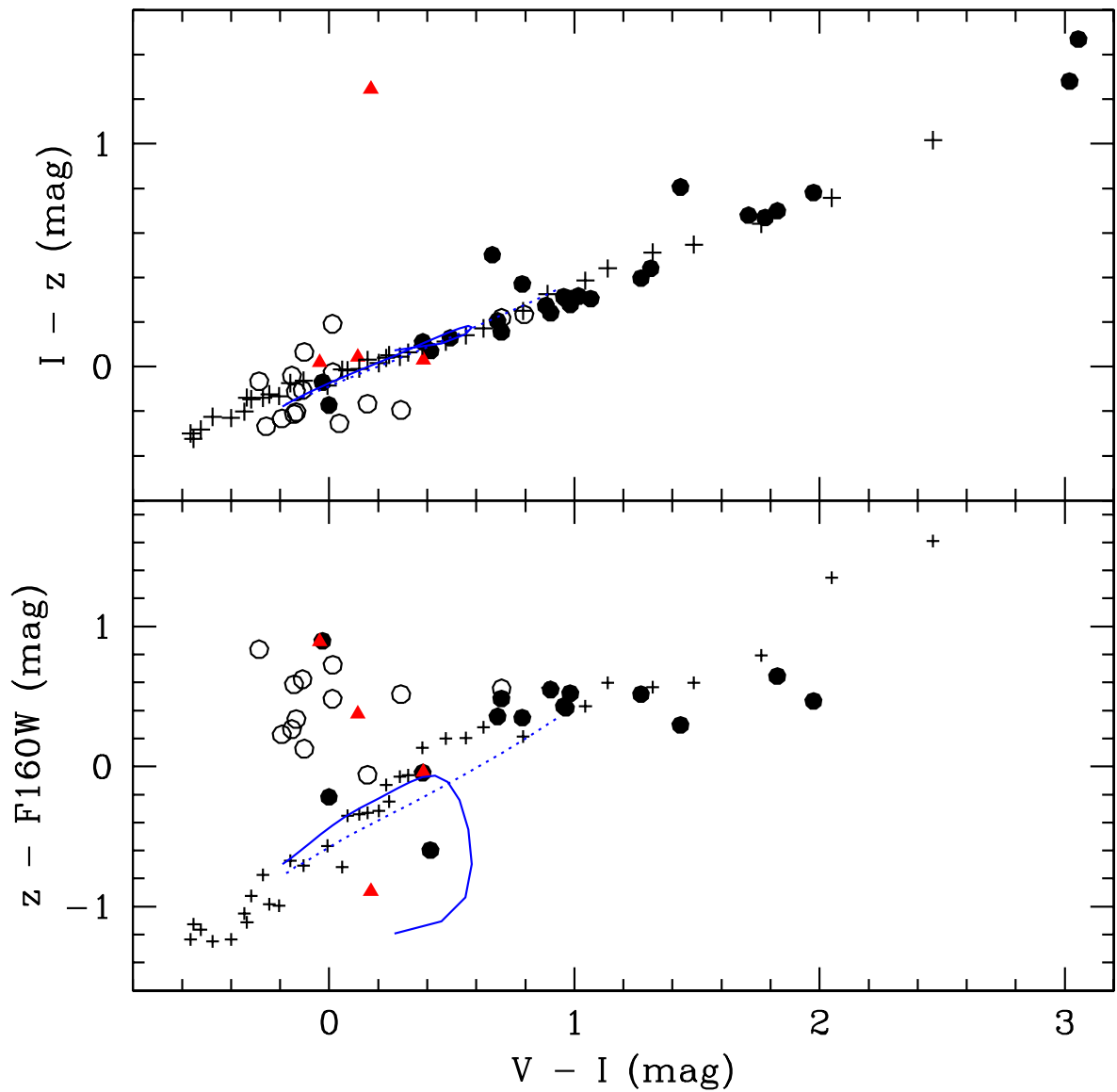}
\caption{Optical and infrared color-color diagrams for the 46 unresolved source candidates in the UDF.
Filled and open circles show bright ($I<27$ mag) and faint ($I>27$ mag) objects, respectively. Triangles mark the
spectroscopically confirmed quasars, and the plus symbols show the synthetic colors of \citet{pickles98} stellar
templates for O5 to M6 type dwarfs. Colors for 3000 - 10,000 K pure H (solid lines) and 3500 - 10,000 K pure He
(dotted lines) atmosphere white dwarf models are also shown.}
\end{figure}

\begin{figure}
\plotone{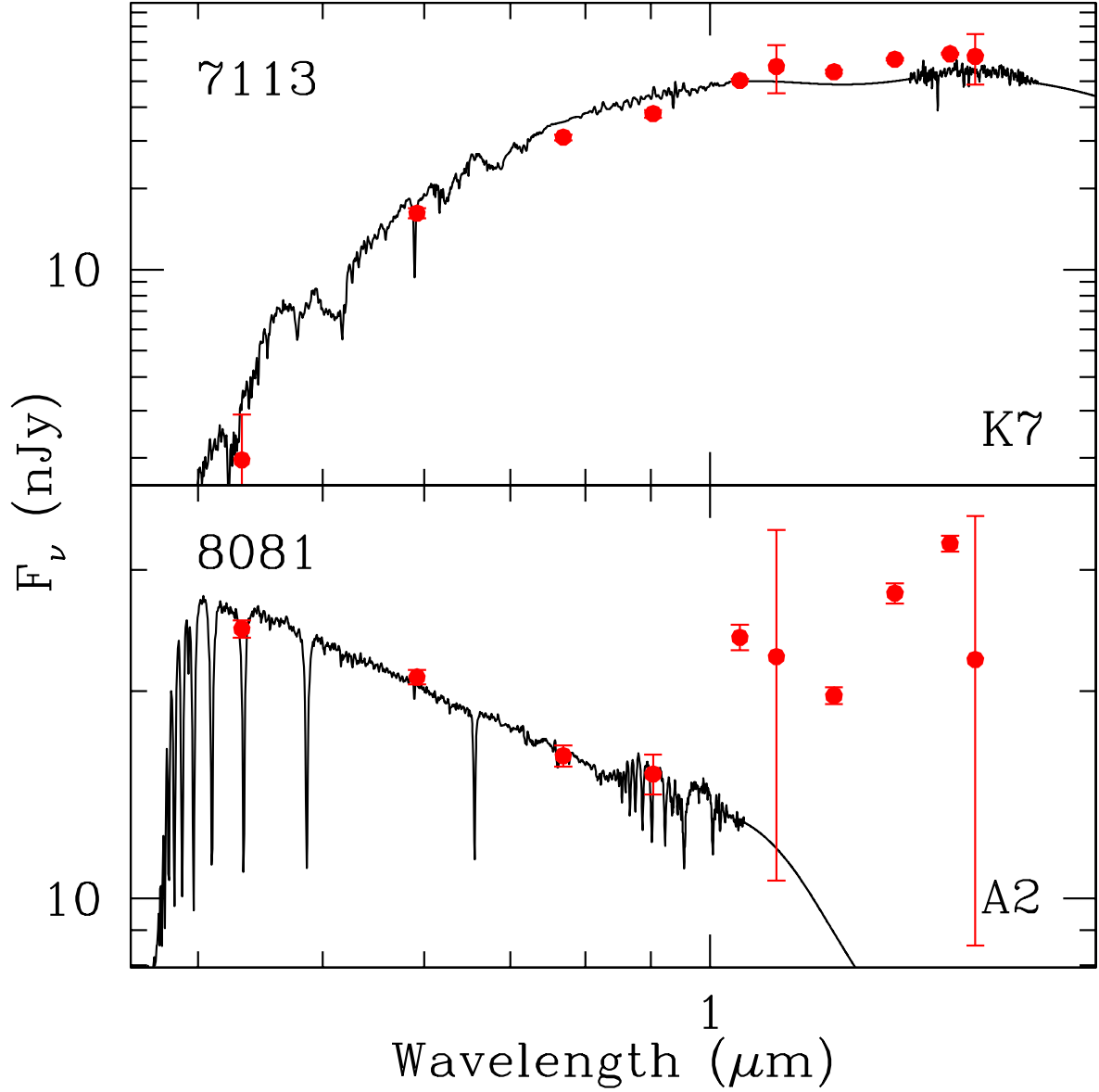}
\caption{Optical and infrared SEDs and the best-fitting main-sequence star templates \citep{pickles98} for the
faint ($I>27$ mag) unresolved sources.}
\end{figure}

\begin{figure}
\plotone{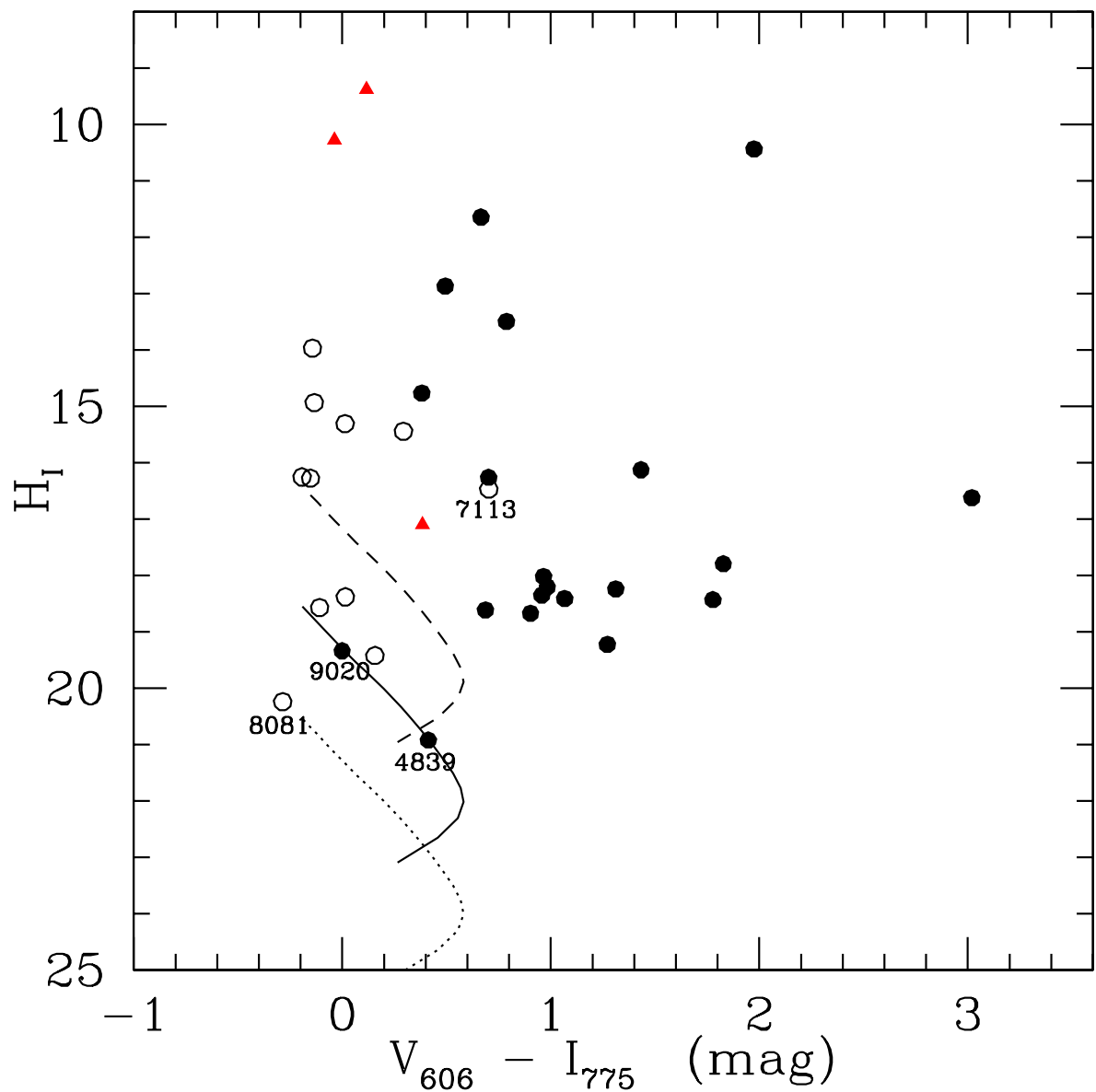}
\caption{Reduced proper motion diagram for the unresolved source candidates in the UDF. Filled and open circles
show bright ($I<27$ mag) and faint ($I>27$mag) sources, respectively. Triangles mark the spectroscopically confirmed
quasars. Dashed, solid, and dotted lines show the cooling sequences for disk ($V_{\rm tan}=30$ \kms), thick disk
(80 \kms), and halo (200 \kms) white dwarfs.}
\end{figure}

\begin{figure}
\plotone{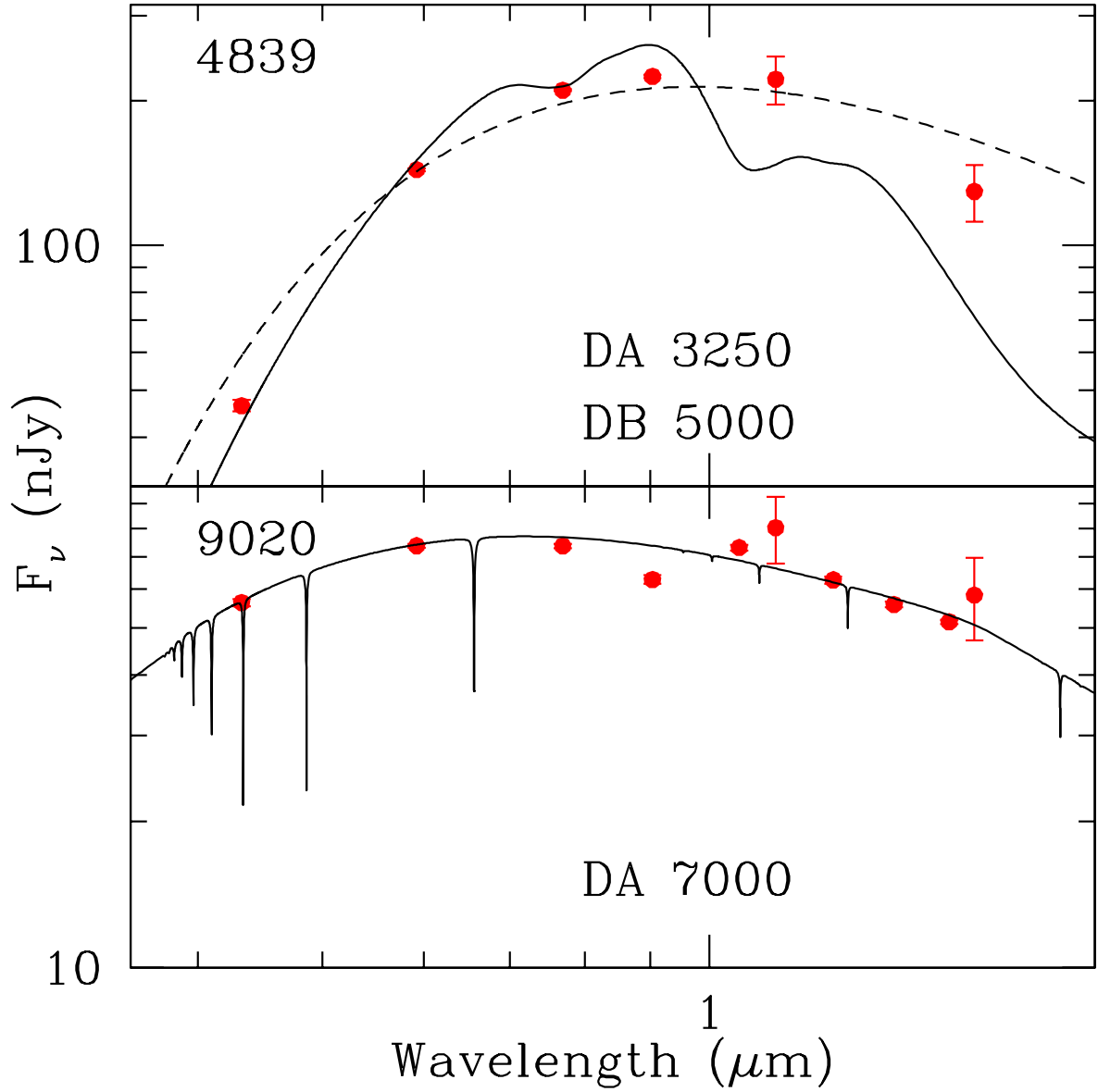}
\caption{Optical and infrared SEDs and the best-fitting pure H (solid lines) and pure He (dashed lines) white dwarf models for UDF 4839 and 9020.}
\end{figure}


\begin{thebibliography}{}
\bibitem[Alcock et al.(2000)]{alcock00} Alcock, C., Allsman, R.~A., Alves, D.~R., et al.\ 2000, \apj, 542, 281 
\bibitem[Beckwith(2005)]{beckwith05} Beckwith, S.~V.~W.\ 2005, VizieR Online Data Catalog, 2258, 0 
\bibitem[Bergeron et al.(2005)]{bergeron05} Bergeron, P., Ruiz, M.~T., Hamuy, M., et al.\ 2005, \apj, 625, 838 
\bibitem[Bergeron et al.(2011)]{bergeron11} Bergeron, P., Wesemael, F., Dufour, P., et al.\ 2011, \apj, 737, 28
\bibitem[Bertin \& Arnouts(1996)]{bertin96} Bertin, E., \& Arnouts, S.\ 1996, \aaps, 117, 393 
\bibitem[Borysow et al.(1997)]{borysow97} Borysow, A., Jorgensen, U.~G., \& Zheng, C.\ 1997, \aap, 324, 185 
\bibitem[Bressan et al.(2012)]{bressan12} Bressan, A., Marigo, P., Girardi, L., et al.\ 2012, \mnras, 427, 127 
\bibitem[Brott \& Hauschildt (2005)]{brott05} Brott, I., \& Hauschildt, P.~H.\ 2005, The Three-Dimensional Universe with Gaia, 576, 565 
\bibitem[Casertano et al.(2000)]{casertano00} Casertano, S., de Mello, D., Dickinson, M., et al.\ 2000, \aj, 120, 2747 
\bibitem[Coe et al.(2006)]{coe06} Coe, D., Ben{\'{\i}}tez, N., S{\'a}nchez, S.~F., et al.\ 2006, \aj, 132, 926 
\bibitem[Ellis et al.(2013)]{ellis13} Ellis, R.~S., McLure, R.~J., Dunlop, J.~S., et al.\ 2013, \apjl, 763, L7 
\bibitem[Gilmore et al.(1990)]{gilmore90} Gilmore, G., King, I.~R., van der Kruit, P.~C., \& Buser, R.\ 1990, Science, 250, 703 
\bibitem[Ibata et al.(1999)]{ibata99} Ibata, R.~A., Richer, H.~B., Gilliland, R.~L., \& Scott, D.\ 1999, \apjl, 524, L95 
\bibitem[Koekemoer et al.(2012)]{koekemoer12} Koekemoer, A.~M., Ellis, R.~S, McLure, R.~J., et al.\ 2012, arXiv:1212.1448 
\bibitem[Kilic et al.(2004)]{kilic04} Kilic, M., von Hippel, T., Mendez, R.~A., \& Winget, D.~E.\ 2004, \apj, 609, 766 
\bibitem[Kilic et al.(2005)]{kilic05} Kilic, M., Mendez, R.~A., von Hippel, T., \& Winget, D.~E.\ 2005, \apj, 633, 1126 
\bibitem[Kilic et al.(2006)]{kilic06} Kilic, M., Munn, J.~A., Harris, H.~C., et al.\ 2006, \aj, 131, 582 
\bibitem[Kilic et al.(2010)]{kilic10} Kilic, M., Leggett, S.~K., Tremblay, P.-E., et al.\ 2010, \apjs, 190, 77 
\bibitem[Kowalski \& Saumon(2006)]{kowalski06} Kowalski, P.~M., \& Saumon, D.\ 2006, \apjl, 651, L137 
\bibitem[Mahmud \& Anderson(2008)]{mahmud08} Mahmud, N., \& Anderson, J.\ 2008, \pasp, 120, 907 
\bibitem[M{\'e}ndez \& Minniti(2000)]{mendez00} M{\'e}ndez, R.~A., \& Minniti, D.\ 2000, \apj, 529, 911 
\bibitem[Munn et al.(2004)]{munn04} Munn, J.~A., Monet, D.~G., Levine, S.~E., et al.\ 2004, \aj, 127, 3034 
\bibitem[Pham(1997)]{pham97} Pham, H.-A.\ 1997, Hipparcos - Venice '97, 402, 559 
\bibitem[Pickles(1998)]{pickles98} Pickles, A.~J.\ 1998, \pasp, 110, 863 
\bibitem[Pirzkal et al.(2004)]{pirzkal04} Pirzkal, N., Xu, C., Malhotra, S., et al.\ 2004, \apjs, 154, 501 
\bibitem[Pirzkal et al.(2005)]{pirzkal05} Pirzkal, N., Sahu, K.~C., Burgasser, A., et al.\ 2005, \apj, 622, 319 
\bibitem[Reid \& Majewski(1993)]{reid93} Reid, N., \& Majewski, S.~R.\ 1993, \apj, 409, 635 
\bibitem[Reid(2005)]{reid05} Reid, I.~N.\ 2005, \araa, 43, 247 
\bibitem[Rowell \& Hambly(2011)]{rowell11} Rowell, N., \& Hambly, N.~C.\ 2011, \mnras, 417, 93 
\bibitem[Schmidt(1968)]{schmidt68} Schmidt, M.\ 1968, \apj, 151, 393 
\bibitem[Sion et al.(2009)]{sion09} Sion, E.~M., Holberg, J.~B., Oswalt, T.~D., McCook, G.~P., \& Wasatonic, R.\ 2009, \aj, 138, 1681 
\bibitem[Tremblay \& Bergeron(2009)]{tremblay09} Tremblay, P.-E., \& Bergeron, P.\ 2009, \apj, 696, 1755 
\bibitem[von Hippel \& Bothun(1990)]{vonhippel90} von Hippel, T., \& Bothun, G.\ 1990, \aj, 100, 403 
\bibitem[Williams et al.(1996)]{williams96} Williams, R.~E., Blacker, B., Dickinson, M., et al.\ 1996, \aj, 112, 1335 
\end{thebibliography}
\end{document}